\documentclass[twocolumn,english,superscriptaddress]{revtex4-2}
\usepackage[T1]{fontenc}
\usepackage[latin9]{inputenc}
\usepackage{color}
\usepackage{babel}
\usepackage{units}
\usepackage{amsmath}
\usepackage{amssymb}
\usepackage{stmaryrd}
\usepackage{graphicx}
\usepackage[unicode=true,pdfusetitle,
 bookmarks=true,bookmarksnumbered=false,bookmarksopen=false,
 breaklinks=false,pdfborder={0 0 0},pdfborderstyle={},backref=false,colorlinks=true]
 {hyperref}

\makeatletter

\providecommand{\tabularnewline}{\\}

\makeatother

\begin{document}
\title{Random-strain-induced correlations in materials with intertwined nematic
and magnetic orders}
\author{W. Joe Meese}
\affiliation{School of Physics and Astronomy, University of Minnesota, Minneapolis,
Minnesota 55455, USA}
\author{Thomas Vojta}
\affiliation{Department of Physics, Missouri University of Science and Technology,
Rolla, Missouri 65409, USA}
\author{Rafael M. Fernandes}
\affiliation{School of Physics and Astronomy, University of Minnesota, Minneapolis,
Minnesota 55455, USA}
\begin{abstract}
Electronic nematicity is rarely observed as an isolated instability
of a correlated electron system. Instead, in iron pnictides and in certain
cuprates and heavy-fermion materials, nematicity is intertwined with
an underlying spin-stripe or charge-stripe state. As a result, random
strain, ubiquitous in any real crystal, creates both random-field disorder
for the nematic degrees of freedom and random-bond disorder for the
spin or charge ones. Here, we put forward an Ashkin-Teller model with
random Baxter fields to capture the dual role of random strain in
nematic systems for which nematicity is a composite order arising from a stripe
state. Using Monte Carlo to simulate this \emph{random Baxter-field
model}, we find not only the expected break-up of the system into
nematic domains, but also the emergence of nontrivial disorder-promoted
magnetic correlations. Such correlations enhance and tie up the fluctuations
associated with the two degenerate magnetic stripe states from which
nematicity arises, leaving characteristic signatures in the spatial
profile of the magnetic domains, in the configurational space of the spin variables,
and in the magnetic noise spectrum. We discuss possible experimental
manifestations of these effects in iron-pnictide superconductors.
Our work establishes the random Baxter-field model as a more complete
alternative to the random-field Ising model to describe complex electronic
nematic phenomena in the presence of disorder.
\end{abstract}
\maketitle

\section{Introduction}

Even weak lattice disorder and structural inhomogeneity can substantially
alter the properties of electronic ordered states \citep{Vojta13,vojta2019disorder_condmat_rev}.
For instance, pair-breaking promoted by local impurity scattering
strongly reduces the transition temperature of unconventional superconductors
\citep{BalianWerthamer63,Dalichaouch_etal95,Mackenzie_etal98},
and impurity-induced phase slips can destroy the long-range order in vortex lattices
or incommensurate charge density waves \cite{larkin1970RFIM,Gruner88,GiamarchiLeDoussal94}.
Because of the unavoidable coupling of the nematic order parameter to lattice deformations,
electronic nematicity is particularly sensitive to the intrinsic inhomogeneities of the crystal lattice.
Electronic nematic order is characterized by the spontaneous breaking
of the rotational symmetry of the system due to electron-electron interactions
while the translational symmetry is preserved \citep{Kivelson1998,fernandes2014what_drives}.
For tetragonal systems such as cuprates, heavy-fermion materials,
and iron-based superconductors, the experimentally observed nematic
order has an Ising symmetry, associated with choosing one of the two
axes (or one of the two diagonals) of the square basal plane of the unit cell \citep{Fradkin_review}.
Due to the electron-phonon coupling, the electronic symmetry breaking
is accompanied by a lowering of the tetragonal symmetry of the lattice which
undergoes an orthorhombic
distortion \citep{Qi2009,fernandes2010elasticproperties,Paul2010,chu2012divergent_nematic_susc,Bohmer2014}.
Such a linear coupling between shear strain and the electronic nematic
order parameter renders random strain, which originates from local
lattice defects and impurities, a random (Ising) field for the nematic order
parameter \citep{Fradkin_review}.
As a result, the random-field Ising model (RFIM) has been employed
to model nematic-related phenomena in crystals \citep{carlson_kivelson_2006_hysteresis,loh_dahmen2010noise,Carlson2011,carlson2015decoding}.
A hallmark of this widely studied model is the complete breakup of
the Ising ordered ground state into domains, which takes place for
any random-field strength in two-dimensional (2D) systems and beyond
a critical disorder strength in 3D systems \citep{imry1975random,binder1983random,nattermann1998theory,Zachar2003}.
More specifically, the RFIM has been invoked to explain puzzling experimental
observations in the local density-of-states of underdoped cuprates
\citep{Phillabaum2012}, in the spin-lattice relaxation rate of iron
selenide \citep{schmalian_2021_mft_rfim_prb}, and in the elasto-resistance
of doped iron pnictides \citep{Kuo2016}.

Despite the progress in understanding the interplay between nematicity
and random strain, one important ingredient has been missing. In many
of the tetragonal systems where it has been observed, nematicity is
often intertwined with a density-wave type of order that breaks both
rotational and translational symmetries \citep{Fradkin_Kivelson_Tranquada,fernandes2019intertwined}.
In the cuprates YBa$_{2}$Cu$_{3}$O$_{6+x}$ \citep{Taillefer2015}
and La$_{2-x}$Sr$_{x}$CuO$_{4}$ \citep{Achkar2016}, and in BaNi$_{2}$As$_{2}$
\citep{Eckberg2020} (a relative of the iron-based superconductors),
it is a charge density-wave whereas in the heavy-fermion CeAuSb$_{2}$
\citep{Seo2020} and in the iron-arsenides family \citep{Johnston2010}
(BaFe$_{2}$As$_{2}$, LaFeAsO, and NaFeAs), it is a spin density-wave.
In all cases, the density-wave tends to be stripe-like, characterized
by two degenerate ordering vectors $\boldsymbol{Q}_{1}$ and $\boldsymbol{Q}_{2}$
related by a fourfold rotation. In analogy with classical liquid crystals,
electronic nematicity in these compounds has been proposed to be a
vestigial order \citep{Nie2014,fernandes2019intertwined} (i.e. a
partially melted version) of the underlying density wave -- which
plays the role of an electronic smectic phase \citep{xu2008ising_spin_orders,fang2008theoryofelectronicnematicity,fernandes2010elasticproperties,Mesaros2011,Kamiya2011,Fernandes2012,Nie2014,Yuxuan2014,Nie2017,Orth2019,Mukhopadhyay2019}.
As a result, the nematic order parameter is described as a composite
density-wave order parameter.

The composite nature of the nematic order parameter points to a more
complex role of random strain that goes beyond the breakup of long-range
order into nematic domains. To see this, consider the iron arsenides,
for which there is strong evidence that nematic order is a vestige
of the stripe magnetic ground state \citep{Fernandes2013}. The latter
has wave-vectors $\boldsymbol{Q}_{1}=(\pi,0)$ and $\boldsymbol{Q}_{2}=(0,\pi)$
and is described by the staggered magnetizations $\boldsymbol{M}_{A}$
and $\boldsymbol{M}_{B}$ of the two sublattices of the square lattice,
which can be either parallel or anti-parallel to each other (see Fig.
\ref{fig:J1-J2_ground_states}). The Ising-nematic order parameter
$\phi$ in this case is the composite $\phi=\boldsymbol{M}_{A}\cdot\boldsymbol{M}_{B}$
\citep{chadra1990j1j2model}. Random shear strain $\epsilon$ therefore
plays a dual role. On the one hand, it leads to randomness of the magnetic
interactions and thus to random-bond disorder for $\boldsymbol{M}_{i}$.
On the other hand, it couples linearly to $\phi$ and therefore creates
random-field disorder for the nematic order parameter. Because random-field and random-bond
effects are quite different, and because there is mutual feedback
between the coupled nematic and magnetic degrees of freedom, a description
in terms of the RFIM is incomplete.

In this paper, we propose and solve an effective model that, like
the RFIM, requires no prior knowledge about microscopic details of
the system but, unlike the RFIM, accounts for the dual random-field/random-bond
role of random strain in nematic systems for which nematicity is a
composite order, rather than an isolated instability. It consists
of the Ashkin-Teller Hamiltonian in the presence of a random Baxter
field, and is thus dubbed the \emph{random Baxter-field model }(RBFM):

\begin{equation}
H=-\sum_{\left\langle ij\right\rangle }\left[J\left(\sigma_{i}\sigma_{j}+\tau_{i}\tau_{j}\right)+K\sigma_{i}\tau_{i}\sigma_{j}\tau_{j}\right]-\sum_{i}\varepsilon_{i}\sigma_{i}\tau_{i}.\label{H_RBFM}
\end{equation}

Although the model is based on the mapping of the Ising $J_{1}$-$J_{2}$
model onto the Ashkin-Teller model \citep{jin_sandvik2012prl,jin_sandvik2013prb},
it should be relevant not only for the case where the primary instability
is a magnetic stripe state with easy-axis anisotropy, but also when
the primary phase is a commensurate charge-stripe phase. Here, the
$\sigma$ and $\tau$ Ising variables describe the fourfold degenerate
ground state, consisting of spin-up and spin-down stripes (or, equivalently,
charge-rich and charge-poor stripes) aligned along the $x$ and $y$
axes (see Fig. \ref{fig:J1-J2_ground_states}). The composite nematic
order paramer is given by $\phi=\sigma\tau$, and couples linearly
with the shear strain $\varepsilon$. The two energy scales $J$ and
$K$ can be directly connected to the $J_{1}$ and $J_{2}$ exchange
constants of the original $J_{1}$-$J_{2}$ model. We note that this
approach is different from and complementary to previous ones where
random-field disorder was included only on the primary charge-stripe
order \citep{Nie2014}.

We use a combination of Metropolis and Replica-Exchange Wang-Landau
(WL) Monte Carlo methods to simulate the RBFM on a square lattice.
Since both cuprates and iron pnictides are layered systems with relatively
large charge-order/magnetic anisotropies, such an approximation is
reasonable. Our results not only show a break-up of the system into
nematic domains, as in the RFIM, but they also reveal unusual magnetic
correlations. Importantly, such correlations are completely absent
in the clean case, and arise entirely from the last term in (\ref{H_RBFM}),
i.e. from the random strain. The key point is that while $\varepsilon_{i}$
completely determines the value of the local nematic order parameter
$\sigma_{i}\tau_{i}$ (either $+1$ or $-1$), it allows two different
sets $\left\{ \sigma_{i},\tau_{i}\right\} $ of magnetic order parameters
(either $\{+1,+1\}$ and $\{-1,-1\}$ or $\{+1,-1\}$ and $\{-1,+1\}$).
Consequently, a particular random-strain realization allows multiple
$\left\{ \sigma_{i},\tau_{i}\right\} $ magnetic configurations --
in contrast, the RFIM has a single ground state for a given random-field
realization.

The consequences of the random-strain driven magnetic correlations
are multifold. In real space, they cause the system to also break up
into magnetic domains. However, the typical sizes of the magnetic
domains are larger than that of the nematic domains by a factor of
order $2$. The disorder-induced magnetic correlations also leave
pronounced signatures in the joint $\left\{ \sigma,\tau\right\} $
distribution in configurational space.
In the clean case, as the temperature is lowered, the shape
of the typical $\left\{ \sigma,\tau\right\} $ distribution changes from
a circle centered at the origin to four sharp peaks at the vertices
of a square. In contrast, in the disordered case, there is a wide
intermediate temperature range for which the typical $\left\{ \sigma,\tau\right\} $
configurations form a hollow square shape, indicative of one Ising
variable acquiring a finite value while the other one explores its
entire configuration axis. Compared to the four sharp $\left\{ \sigma,\tau\right\} $
peaks of the clean case, these enhanced correlations also cause an
unexpected increase of the magnetic susceptibility in the disordered
system. The nematic susceptibility, on the other hand, is always suppressed
with respect to the clean case. This is not the only effect of random
strain on the magnetic and nematic susceptibility: while their peaks
are coincident in the clean case, they split in the disordered case.

Moreover, nontrivial magnetic correlations also appear in time domain.
In the clean case, the two magnetic variables fluctuate in a nearly
independent fashion, and the corresponding power spectrum densities
display a low-frequency plateau. On the other hand, in the disordered
case, the magnetic fluctuations are correlated, in that if one of
the Ising variables fluctuates around a non-zero value, the other
one must fluctuate around zero. This results in coherent switching
events and in power spectrum densities that show no plateau behavior
up to the very low frequencies probed in our simulations. Finally,
we discuss possible experimental manifestations of the various random-strain
induced effects on the magnetic degrees of freedom.

The structure of this paper is as follows. We first motivate and introduce
the RBFM in Section \ref{sec:RBFATM}, as well as discuss some of
its qualitative properties in two dimensions. In Section \ref{sec:Thermodynamics-of-the-RBFATM},
we introduce our main Replica-Exchange WL simulation and discuss the
disorder-averaged thermodynamic behavior seen in the RBFM. Additionally,
we compare our results for the RBFM to our simulations of the RFIM.
Following this, in Section \ref{sec:Dynamics-at-zero-Baxter-coupling},
we discuss the time-dependence of the fluctuations in the RBFM when
we impose relaxational dynamics on the system. Finally, we summarize
and discuss the implications of our results in Section \ref{sec:Discussion}.
Appendix A presents technical details of our Monte Carlo simulations.

\section{Random Baxter Field Ashkin-Teller model (RBFM)\label{sec:RBFATM}}

\subsection{Connection with the iron pnictides and the $J_{1}$-$J_{2}$ model}

The low-energy properties of the magnetic and nematic phases of the iron pnictides are often described by means
of the $J_{1}$-$J_{2}$ model
\citep{xu2008ising_spin_orders,fang2008theoryofelectronicnematicity,Si_Abrahams,Kamiya2011}.
The $J_{1}$-$J_{2}$
Hamiltonian is given by
\begin{equation}
H=J_{1}\sum_{\left\langle ij\right\rangle }\boldsymbol{S}_{i}\cdot\boldsymbol{S}_{j}+J_{2}\sum_{\langle\!\langle ij\rangle\!\rangle}\boldsymbol{S}_{i}\cdot\boldsymbol{S}_{j},\label{eq:j1-j2_model}
\end{equation}
where $\left\langle ij\right\rangle $ and $\langle\!\langle ij\rangle\!\rangle$
denote nearest-neighbor and next-nearest-neighboring sites $i$ and
$j$, respectively. The spins, $\{\boldsymbol{S}_{i}\}$, are classical
${\rm O}(n)$ vectors that sit on a square lattice and interact with
each other via the nearest-neighbor and next-nearest-neighbor exchange
interactions $J_{1}$ and $J_{2}$.

Because the iron pnictide compounds are metallic, the itinerant electrons play
an important role, which is not captured by the $J_{1}$-$J_{2}$
model \citep{Fernandes2012}. Nevertheless, the $J_{1}$-$J_{2}$
model can still be used as an effective low-energy model to gain insight
into the magnetic and nematic properties of the iron pnictides. Indeed,
the phase diagram of this model contains both a stripe magnetic phase
and a vestigial nematic phase \citep{chadra1990j1j2model}.

\begin{figure}
\includegraphics[width=1\columnwidth]{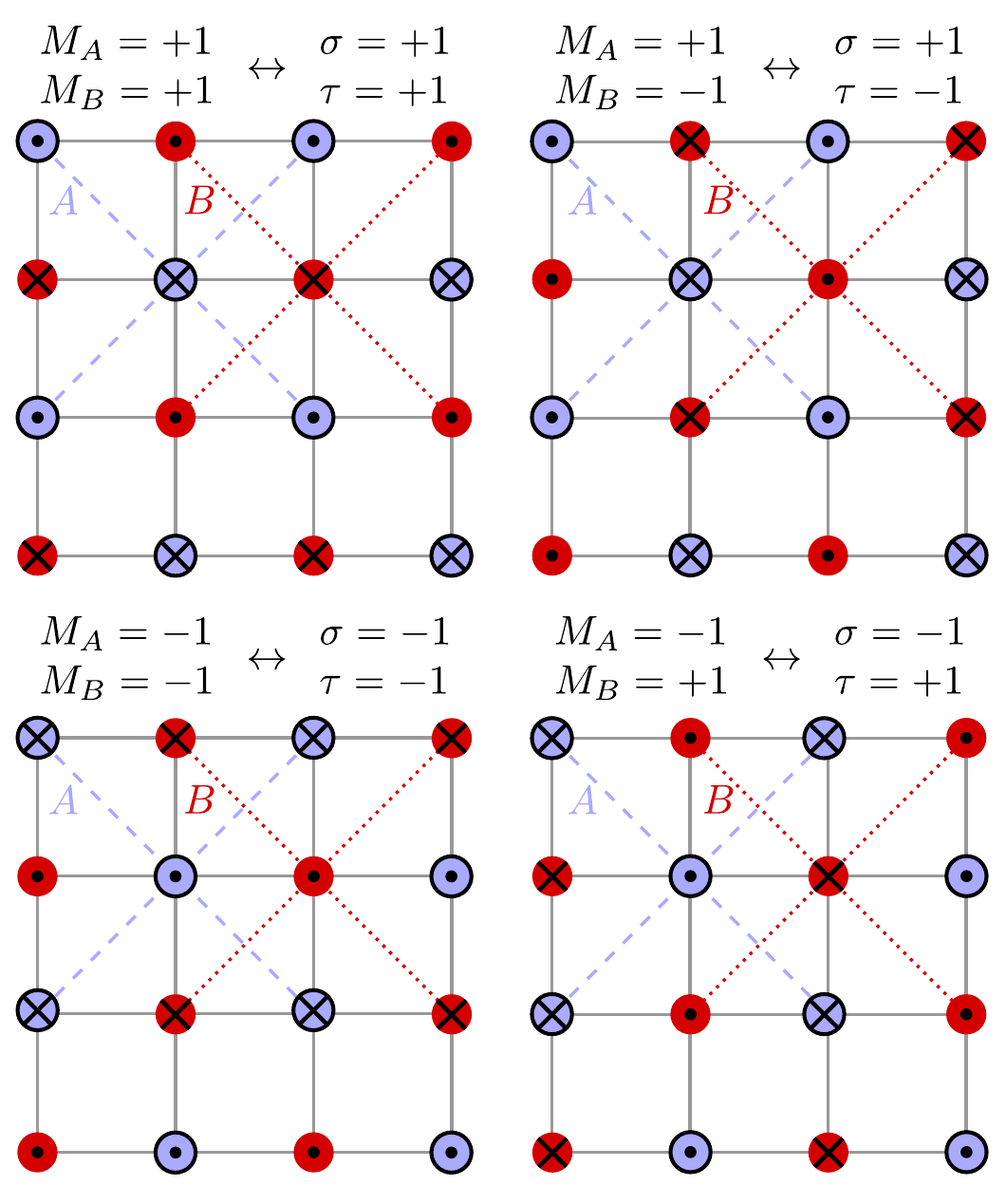}

\caption{The fourfold degenerate stripe-magnetic ground state of the Ising
$J_{1}$-$J_{2}$ model on a square lattice. Blue and red sites denote
the $A$ and $B$ sublattices. The magnetic moments point out of the
plane. The horizontal (vertical) stripe configuration, corresponding
to the ordering vector $\boldsymbol{Q}_{H}=\left(0,\pi\right)$ $\left(\boldsymbol{Q}_{V}=\left(\pi,0\right)\right)$,
corresponds to parallel (anti-parallel) staggered sublattice magnetizations
$M_{A}$ and $M_{B}$. These states exhibit either a positive or negative
nematic order parameter, defined as the product $\phi\propto M_{A}M_{B}$.
The staggered magnetizations on the $A$ and $B$ sublattices of the
Ising $J_{1}$-$J_{2}$ model are in one-to-one correspondence with
the $\sigma$ and $\tau$ Ising variables of the Ashkin-Teller model.
Similarly, the nematic order parameter in the former maps onto the
Baxter variable $\sigma\tau$ of the latter. \label{fig:J1-J2_ground_states}}
\end{figure}

While most studies in the pnictides have focused on the Heisenberg
case $\left(n=3\right)$, the spins in these materials are typically
confined to a single axis due to spin-orbit coupling \citep{dai2015AFM_order_review,christensen2015spin_orbit_coupling}.
Thus, hereafter we focus on the Ising limit $\left(n=1\right)$. The
fourfold-degenerate ground state for the Ising case, shown in Fig.
\ref{fig:J1-J2_ground_states}, consists of spin-up and spin-down
stripes oriented parallel to either the $x$ axis or the $y$ axis.

To understand this degeneracy, we note that in the $J_{1}$-$J_{2}$ model applied to iron pnictides,
the next-nearest-neighbor exchange $J_{2}$ is always antiferromagnetic
while the nearest-neighbor exchange $J_{1}$ can be generally
antiferromagnetic or ferromagnetic,  When $J_{2}/\left|J_{1}\right|=0$,
the ground state is either N\'eel antiferromagnetic or ferromagnetic
depending on whether $J_{1}>0$ or $J_{1}<0$, respectively. In the
opposite limit, when $J_{2}/\left|J_{1}\right|\rightarrow\infty$, the ground
state is stripe antiferromagnetic as the next-nearest-neighboring
spins favor antiferromagnetic alignment, rendering the nearest-neighbors
on the two interpenetrating sublattices shown in Fig. \ref{fig:J1-J2_ground_states}
independent of one another \footnote{While in the Ising case the spins are collinear by construction, for
${\rm O}(n>1)$ realizations of the $J_{1}$-$J_{2}$ model, fluctuations
give rise to a biquadratic term in the Hamiltonian that lock the staggered
magnetizations of the two sublattices to be collinear \citep{chadra1990j1j2model},
giving rise to a stripe magnetic state similar to that in Fig. \ref{fig:J1-J2_ground_states}.}.
When $J_{2}$ and $\left|J_{1}\right|$ are comparable, there is competition between
the $J_{1}$ and $J_{2}$ terms, as the $J_{1}$ term favors \emph{ferromagnetic}
alignment of next-nearest-neighboring spins, whereas the $J_{2}$
term prefers these neighbors to have \emph{antiferromagnetic} alignment.
For this reason, the model is said to be frustrated.

The transition from the N\'eel or ferromagnetic ground state to the stripe-ordered ground state
occurs at $J_{2}/\left|J_{1}\right|=1/2$. We therefore focus on the range $J_{2}/\left|J_{1}\right|>1/2$.
In this regime, the Hamiltonian in Eq. (\ref{fig:J1-J2_ground_states})
can be written as two antiferromagnetic Ising models on the $A$ and
$B$ sublattices of the square lattice coupled by the $J_{1}$ term.
Therefore, the stripe-magnetic order parameter has two components
comprised of the staggered sublattice magnetizations, given by
\begin{align}
M_{A,B} & =\sum_{\left(i_{x},i_{y}\right)\in A,B}^{N_{S}}\left(-1\right)^{i_{x}+i_{y}}S_{i}.\label{eq:MAB_definition}
\end{align}
In the equation above, the summation runs over $\left(i_{x},i_{y}\right)$
integer pairs that index the sites that lie in either the $A$ or
$B$ sublattice, each containing half of the total number of spins
$\left(N_{S}=N/2\right)$.

The nematic order parameter, $\phi\propto M_{A}M_{B}$, is composite
in terms of the primary magnetic order parameters $M_{i}$. It is
Ising-nematic because the sublattice magnetizations can be either
aligned $\left(\phi>0\right)$ or anti-aligned $\left(\phi<0\right)$.
These two cases represent the horizontal $\left(\phi>0\right)$ and
vertical $\left(\phi<0\right)$ stripe configurations, as shown in
Fig. \ref{fig:J1-J2_ground_states}. In the Ising $J_{1}$-$J_{2}$
model \citep{jin_sandvik2013prb}, long-range nematic order appears
simultaneously with the primary stripe magnetic phase, i.e.,
there is no separate nematic phase \footnote{In the $J_{1}$-$J_{2}$ model with classical vector spins, there
is generally a nematic phase between the high-temperature paramagnetic
phase and the ground state stripe magnetic phase \citep{fang2008theoryofelectronicnematicity,Fernandes2012}.
This phase is characterized as one with $\left\langle \boldsymbol{M}_{A}\cdot\boldsymbol{M}_{B}\right\rangle \neq0$
despite that $\left\langle \boldsymbol{M}_{A}\right\rangle =\left\langle \boldsymbol{M}_{B}\right\rangle =0$.
In purely two-dimensional systems, the stripe magnetic phase only
exists at $T=0$ by the Mermin-Wagner theorem. However, the nematic
order persists at finite temperature since it breaks a discrete symmetry
instead of a continuous one.}.

The stripe configurations lower the tetragonal symmetry of the lattice
to orthorhombic, since bonds along the $x$ axis connect parallel
(anti-parallel) spins, whereas bonds along the $y$ axis connect anti-parallel
(parallel) spins. Due to magneto-elastic coupling, the nematic order
parameter couples bilinearly to strain $\varepsilon$ in the Ginzburg-Landau
free-energy, $\varepsilon\phi\propto\varepsilon\left(M_{A}M_{B}\right)$
\citep{fernandes2010elasticproperties,Paul2010,chu2012divergent_nematic_susc}.
Here, the strain takes on the form
\begin{equation}
\varepsilon\equiv\partial_{x}u_{x}-\partial_{y}u_{y},\label{eq:strain_definition}
\end{equation}
with $\boldsymbol{u}=\left(u_{x},u_{y}\right)$ denoting the lattice
displacement.

\subsection{Mapping onto the Ashkin-Teller model}

The existence of a fourfold degenerate ground state suggests that
the properties of the Ising $J_{1}$-$J_{2}$ model may be captured
by a model with $Z_{4}$ symmetry. That this is indeed the case was
shown in Refs. \citep{jin_sandvik2012prl,jin_sandvik2013prb}, which
numerically determined the phase diagram of the Ising $J_{1}$-$J_{2}$
model. It was found that for $1/2<J_{2}/\left|J_{1}\right|\lesssim0.67$,
there is a first-order transition from a paramagnetic phase into a
stripe phase, whereas for $J_{2}/\left|J_{1}\right|\gtrsim0.67$,
the system undergoes a single second-order phase transition. Moreover,
in the $J_{2}/\left|J_{1}\right|\rightarrow\infty$ limit, the transition
is in the Ising universality class, whereas for $J_{2}/\left|J_{1}\right|\approx0.67$,
the transition has the 4-state Potts universality. These two points
are connected by a line of second-order transitions that displays
\emph{weak universality,} i.e. only the anomalous exponent $\eta=1/4$
is universal while the others depend on the ratio $J_{2}/\left|J_{1}\right|$
\citep{suzuki1974weak_universality,jin_sandvik2012prl,jin_sandvik2013prb}
\footnote{The conjugate field exponent $\delta$ is also universal for systems that exhibit weak universality, but one can think of it as being fixed 
by its relation to the anomalous exponent through hyperscaling. }. This is the same behavior displayed by the 2D Ashkin-Teller model
(ATM), which is defined by:

\begin{equation}
H_{0}=-\sum_{\left\langle ij\right\rangle }\left[J\left(\sigma_{i}\sigma_{j}+\tau_{i}\tau_{j}\right)+K\sigma_{i}\tau_{i}\sigma_{j}\tau_{j}\right].\label{eq:clean_AT_model}
\end{equation}
In the expression above, there are two Ising spins at every site $i$,
denoted by $\sigma_{i}$ and $\tau_{i}$. These spins are also known
as the magnetic ``colors'' of the model. The Ising exchange $J$
is ferromagnetic and the strength of the Baxter exchange coupling
$K$ represents the degree to which the two Ising magnetic colors
are correlated. When $K=0$, the model corresponds to two copies of
the Ising model (like the Ising $J_{1}$-$J_{2}$ model when $J_{2}/\left|J_{1}\right|\rightarrow\infty$)
whereas when $K=J$, it reduces to the 4-state Potts model (like the
Ising $J_{1}$-$J_{2}$ model when $J_{2}/\left|J_{1}\right|\approx0.67$).

\begin{figure}
\includegraphics[width=1\columnwidth]{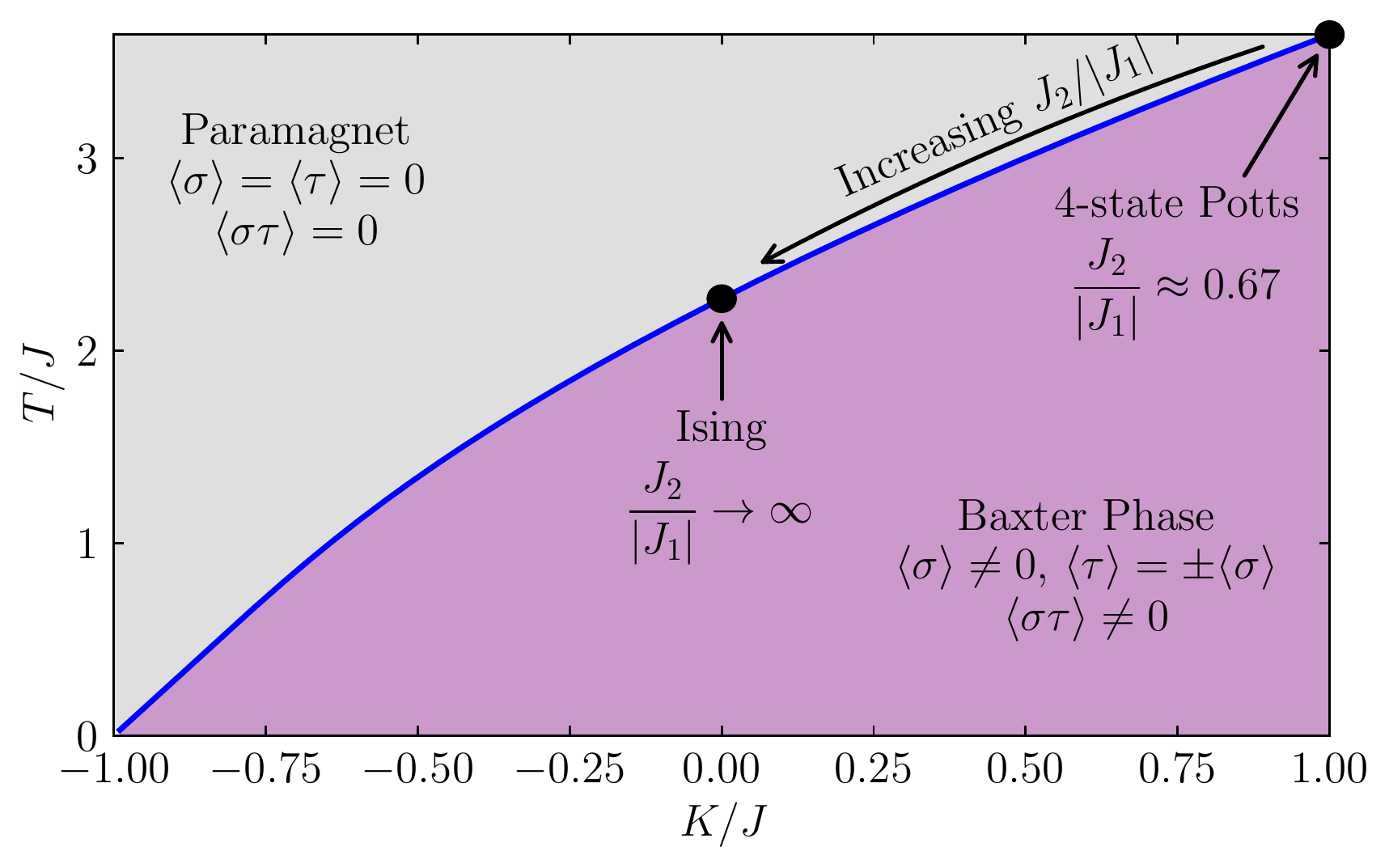}

\caption{Two-dimensional phase diagram of the Ashkin-Teller model along the
temperature $T$ and Baxter coupling $K$ axes, according to Ref.
\citep{kadanoff1980ashkin3d}. A critical line separates the paramagnetic
phase from the Baxter phase, along which only one critical exponent
is universal. The points displaying Ising $(K=0)$ and 4-state Potts
$(K=J)$ universality are indicated, as is the mapping of the Ising
$J_{1}$-$J_{2}$ model with $J_{2}/\left|J_{1}\right|\gtrsim0.67$
onto the $0\protect\leq K/J\protect\leq1$ region of the phase diagram
established in Refs. \citep{jin_sandvik2012prl,jin_sandvik2013prb}.
\label{fig:Baxter_phase_diagram}}
\end{figure}

In fact, the results of Refs. \citep{jin_sandvik2012prl,jin_sandvik2013prb}
established a one-to-one correspondence between the ATM with $0\leq K/J\leq1$
and the Ising $J_{1}$-$J_{2}$ model with $J_{2}/\left|J_{1}\right|\gtrsim0.67$,
such that $K=0$ correspons to $J_{2}/\left|J_{1}\right|\rightarrow\infty$
and $K=J$, to $J_{2}/\left|J_{1}\right|\approx0.67$. The mapping
takes on the following form:
\begin{equation}
\begin{array}{ccc}
\left\langle M_{A}\right\rangle  & \leftrightarrow & \left\langle \sigma\right\rangle ,\\
\left\langle M_{B}\right\rangle  & \leftrightarrow & \left\langle \tau\right\rangle ,\\
\left\langle \phi\right\rangle  & \leftrightarrow & \left\langle \sigma\tau\right\rangle .
\end{array}\label{eq:j1-j2_to_ATM_mapping}
\end{equation}
and is shown in Fig. \ref{fig:Baxter_phase_diagram} superimposed
to the $(K/J,\,T/J)$ phase diagram of the ATM \citep{kadanoff1980ashkin3d}.
Along the interval $K/J\in(-1,1]$, the ground state of the ATM is
fourfold degenerate, and the system cools from a high temperature
paramagnetic phase, characterized by $\left\langle \sigma\right\rangle =\left\langle \tau\right\rangle =\left\langle \sigma\tau\right\rangle =0$,
into the so-called Baxter phase, which displays ferromagnetic order
in both $\sigma$ and $\tau$ such that $\left\langle \sigma\right\rangle =\pm\left\langle \tau\right\rangle $.
Note that the Baxter phase also displays ferro-like order in the composite
Baxter variable $\left\langle \sigma\tau\right\rangle $, which corresponds
to the nematic order parameter of the Ising $J_{1}$-$J_{2}$ model.
Therefore, in this model, the nematic (i.e. Baxter) and magnetic transitions
happen simultaneously via a second-order transition. This is quite
different from the large-$n$ solution of the ${\rm O}(n)$ $J_{1}$-$J_{2}$
model, in which the two transitions are either split or simultaneous
and first-order \citep{fang2008theoryofelectronicnematicity,Fernandes2012}.
The critical line separating the paramagnetic and Baxter phases is
known exactly \citep{jin_sandvik2013prb}:
\begin{equation}
\sinh\left(\frac{2J}{T_{c}}\right)=\exp\left(-\frac{2K}{T_{c}}\right).\label{eq:clean_phase_transition_temperature}
\end{equation}

As discussed above, along this line, only the anomalous exponent is
universal, $\eta=1/4$, while the values of the other exponents (except
for the exponent $\delta$, given by $\delta=\frac{d+2-\eta}{d-2+\eta}=15$)
depend on the ratio $K/J$. We note that ordered phases exist outside
the interval $K/J\in[-1,1]$\citep{kadanoff1980ashkin3d}. For instance,
the Baxter phase for $K/J>1$ is preceded by a phase in which $\left\langle \sigma\tau\right\rangle \neq0$
while $\left\langle \sigma\right\rangle =\left\langle \tau\right\rangle =0$.
Such a behavior is not observed in the Ising $J_{1}$-$J_{2}$ model.
Hereafter, we focus only on the regime $K/J\in(0,1)$ of the ATM,
as it maps onto the $J_{2}/\left|J_{1}\right|\gtrsim0.67$ regime
of the Ising $J_{1}$-$J_{2}$ model.

\subsection{Random strain and the emergence of two length scales}

The role that structural disorder plays on the intertwined magnetic
and nematic phases represents a new type of quenched disorder in statistical
physics. Random strains in the lattice appear as a random-field disorder
for the composite nematic degrees of freedom, whereas they appear
as a random-bond disorder for the primary magnetic degrees of freedom.
Typically, random-field disorder is more acute than random-bond disorder.
Indeed, even a weak random field locally breaks symmetries of the
Hamiltonian and can lead to domain break-up at $T=0$ \citep{imry1975random}.
On the other hand,
the system's exchange constant may overcome the effects of a weak
bond disorder \citep{Vojta13,vojta2019disorder_condmat_rev,vojta2006rare_region_effects_review}.
Random-strain disorder is different than either the pure random-field
or the pure random-bond disorder because the interplay between the
magnetic and nematic degrees of freedom allow for the two types of
disorder effects to feedback onto each other.

Random-strain disorder, denoted by the local variable $\varepsilon_{i}$,
appears as an on-site term through the bilinear coupling to the Ising-nematic
variable of the problem. For an isolated Ising-nematic instability,
as discussed in Refs. \citep{carlson_kivelson_2006_hysteresis,loh_dahmen2010noise,Carlson2011,carlson2015decoding},
this results in the random-field Ising-model (RFIM). In the case of
the ATM considered here, random-strain disorder maps onto a \emph{random
Baxter field}, as the nematic order parameter is the composite $\sigma\tau$.
This leads us to the \emph{random Baxter field Ashkin-Teller model}
(RBFM), presented in Eq. (\ref{H_RBFM}) and repeated here for convenience:
\begin{equation}
H=-\sum_{\left\langle ij\right\rangle }\left[J\left(\sigma_{i}\sigma_{j}+\tau_{i}\tau_{j}\right)+K\sigma_{i}\tau_{i}\sigma_{j}\tau_{j}\right]-\sum_{i}\varepsilon_{i}\sigma_{i}\tau_{i}.\label{eq:RBFATM_Hamiltonian}
\end{equation}
In our simulations, disorder is taken to be spatially uncorrelated
and sampled from a box distribution:
\begin{equation}
\rho(\varepsilon_{i})=\begin{cases}
\frac{1}{4\varepsilon}, & \varepsilon_{i}\in\left[-2\varepsilon,2\varepsilon\right]\\
0, & {\rm otherwise}
\end{cases},\label{eq:random_field_distribution}
\end{equation}

Consequently, $\varepsilon$ denotes the typical disorder strength.
We note that while the disorder term in Eq. (\ref{eq:RBFATM_Hamiltonian})
acts only as a random-field for the Baxter variable $\sigma\tau$, a random-bond term is generated 
by fluctuations (e.g. in the renormalization-group flow) for the individual magnetic variables $\sigma$ and $\tau$. Thus, the RBFM captures the feedback
that exists between the random-field and random-bond effects promoted
by random-strain disorder in the iron pnictides.

\begin{figure*}
\includegraphics[width=1\textwidth]{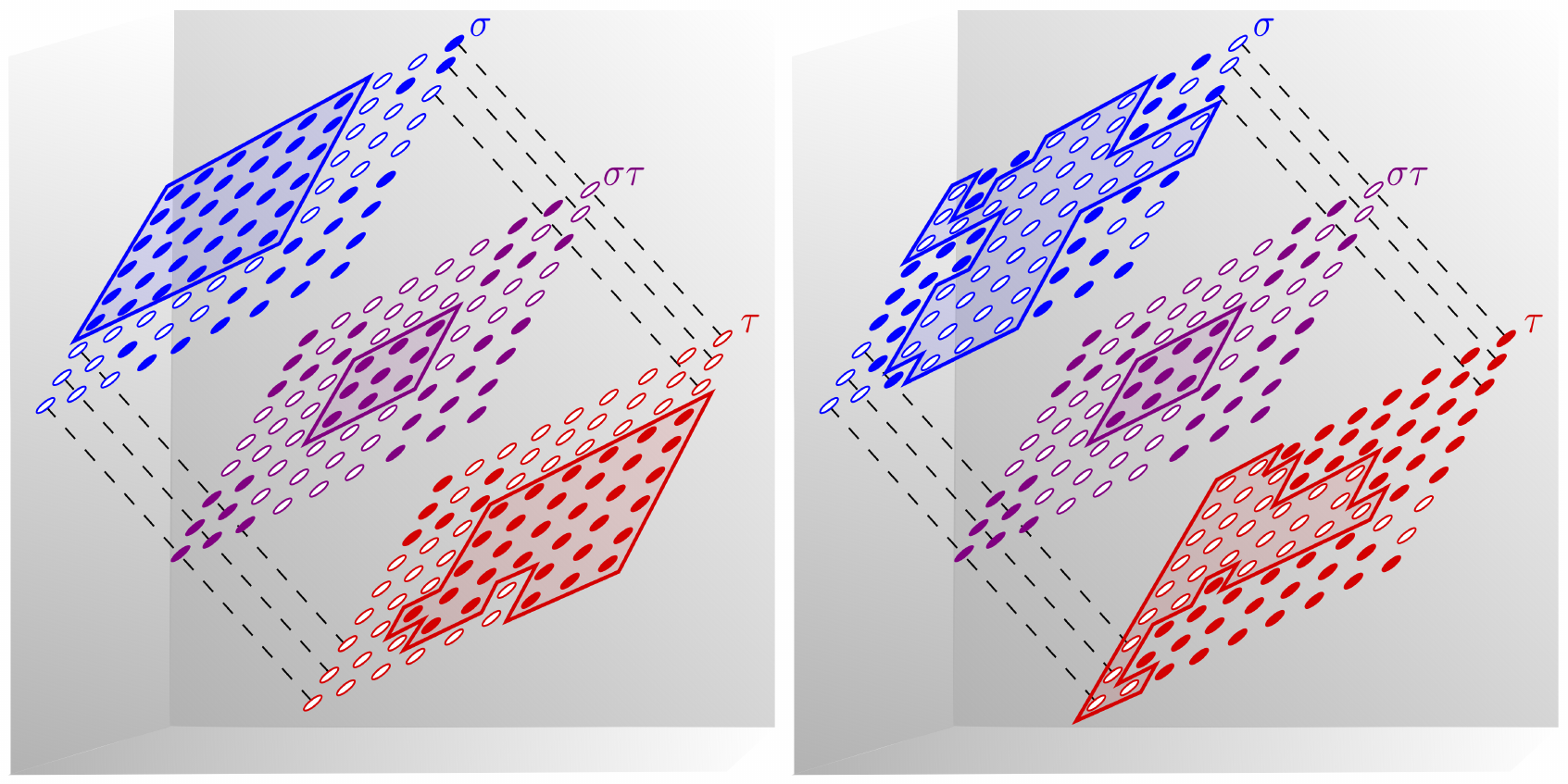}

\caption{Each random Baxter field configuration fixes a Baxter domain configuration,
as shown in the central view of the two dimensional lattice in this
sketch (purple). In both panels, the Baxter ($\sigma\tau$) domain
configuration is identical, but the magnetic domains $\sigma$ (red)
and $\tau$ (blue) are different, as shown by the top and bottom views,
respectively. \label{fig:domain_decomposition_sketch}}
\end{figure*}

\begin{figure*}
\includegraphics[width=1\textwidth]{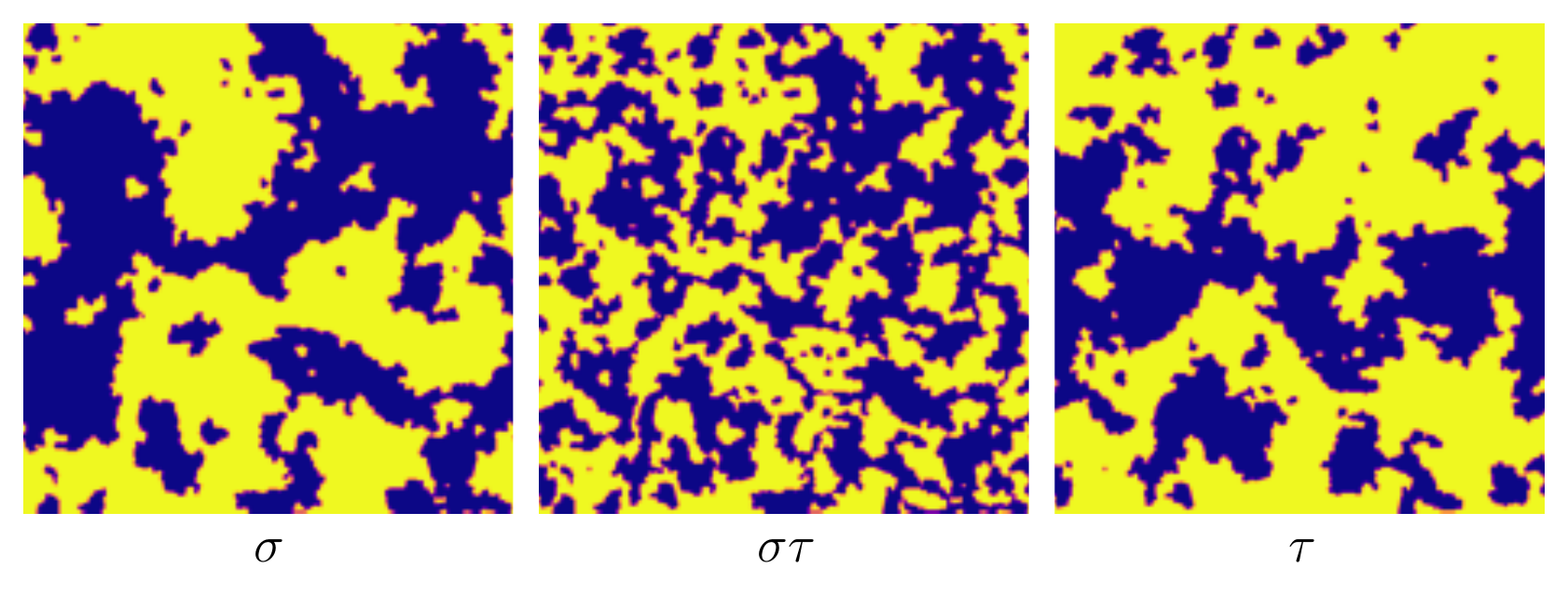}

\caption{Low-temperature domain structure generated by a simulated annealing
run of Eq. (\ref{eq:RBFATM_Hamiltonian}) for a single disorder configuration
with $K=0$, $\varepsilon=J$, and $L=400$. The Baxter domains in
the $\sigma\tau$ panel are seen to be typically smaller than the
magnetic domains in the $\sigma$ and $\tau$ panels. The yellow (light)
regions are those for which the field values are $+1$, whereas the
blue (dark) regions correspond to $-1$. The red lines highlight the
domain walls. \label{fig:domain_decomposition_simulated_annealing}}
\end{figure*}

\begin{figure}
\includegraphics[width=1\columnwidth]{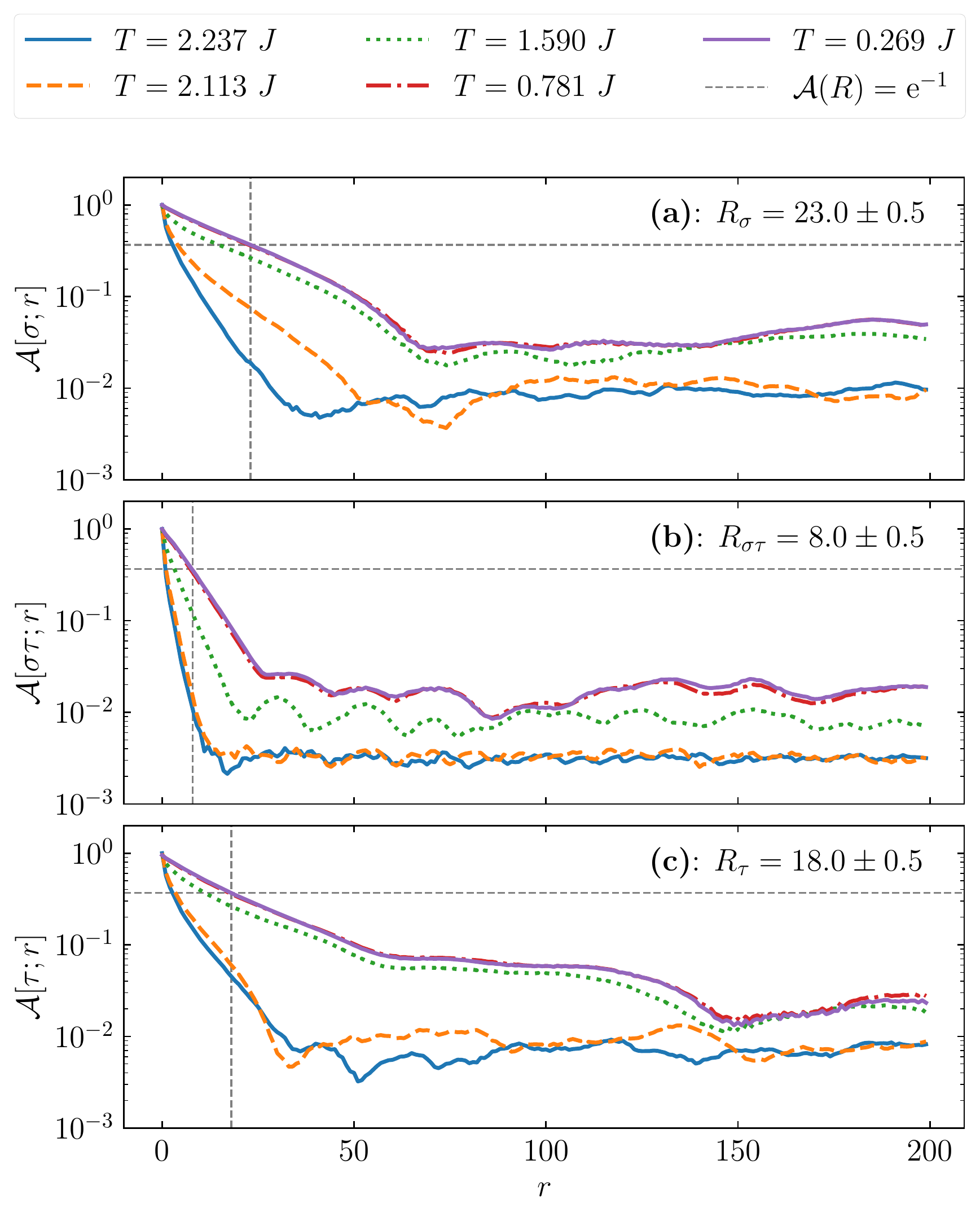}

\caption{Isotropic spin autocorrelation functions $\mathcal{A}(r)$ of the
\textbf{(a)} $\sigma$, \textbf{(b)} Baxter ($\sigma\tau$), and \textbf{(c)}
$\tau$ variables as functions of the radial coordinate $r\equiv\sqrt{x^{2}+y^{2}}$
at various temperatures during a simulated annealing regimen of a
single disorder configuration. The vertical lines are positioned at
the radius $R$ where $\mathcal{A}(R)={\rm e}^{-1}$ at the lowest
temperature probed. \label{fig:Isotropic-spin-spin-correlation_functions}}
\end{figure}

Following the Imry-Ma-Binder analysis \citep{imry1975random,binder1983random},
the Baxter phase of the clean 2D ATM, subject to the random Baxter
field in Eq. (\ref{eq:RBFATM_Hamiltonian}), will break apart into
domains even at $T=0$. The length scale at which this occurs is given
by the breakup length
\begin{equation}
\ell_{b}\sim\exp\left[C\left(\frac{J+K}{\varepsilon}\right)^{2}\right],\label{eq:breakup_length}
\end{equation}
that grows exponentially with the ratio between the ``total'' exchange
and the disorder strength. In the expression above, $C$ is a constant
of order one. Thus, we do not expect long-range order at any temperature
in the thermodynamic limit.

Because the random Baxter field acts on an Ising composite variable,
the domain breakup occurs specifically for the composite variable.
This means domain walls form when the product $\sigma\tau$ changes
value. This is achieved when either $\sigma$ or $\tau$ changes value,
\emph{but not if both do. }At first sight, this seems to imply that
one magnetic order parameter could maintain long-range order while
the other one breaks up to satisfy the constraints of the random Baxter
field. However, such a state is degenerate with one where both orders
eventually break apart, resulting in magnetic domain breakup as well.
Thus, all long-range order is lost in the RBFM at $T=0$.

What makes this model display a richer and more complex behavior than
the RFIM is the fact that disorder introduces nontrivial correlations
between $\sigma$ and $\tau$. Consider, for example, the sizes of
the magnetic domains compared to the nematic domains. In the RFIM,
a single disorder configuration should fix the ground state configuration
\citep{aizenman_wehr1989Gibbs_State}. In the RBFM, meanwhile, because
only one of the two magnetic variables can change at a time across
a nematic domain wall, the typical breakup size of the magnetic domains
is larger than the nematic domains. This introduces a second break-up
length scale into the problem, which is not seen in the RFIM.

Additionally, there is a potentially large degeneracy associated with
the number of ways one can satisfy the disorder constraint with multiple
magnetic domain configurations. This is sketched in Fig. \ref{fig:domain_decomposition_sketch},
where we schematically show identical Baxter domain breakup associated
two different $\sigma$ and $\tau$ domain configurations. Therefore,
these two different magnetic states are degenerate. The residual degeneracy
expected in the ground state of the RBFM contrasts with that of the
RFIM, which is expected to be unique \citep{aizenman_wehr1989Gibbs_State}.
This ground state degeneracy is attributed to the fact that, because
disorder locally favors a value of the composite Baxter variable,
it leaves intact a twofold symmetry of a system.

To assess the difference in domain breakup sizes more clearly, we
show the results of an adaptive simulated annealing run of Eq. (\ref{eq:RBFATM_Hamiltonian})
for a system of linear size $L=400$ with a final temperature of $T=0.269\,J$,
Baxter coupling $K=0$, and a single disorder configuration with
strength $\varepsilon=J$. Details of our simulated annealing regimen
are given in Appendix \ref{subsec:simulated_annealing}. The final
spin values are plotted in Fig. \ref{fig:domain_decomposition_simulated_annealing},
with the yellow (light) and blue (dark) regions representing Ising
domains with value $\pm1$. It is clear that the typical nematic domain
size is smaller than the typical magnetic domain size.

To make a more quantitative comparison, the spin autocorrelation functions
$\mathcal{A}$ for various temperatures during the simulated annealing
regimen are shown in Fig. \ref{fig:Isotropic-spin-spin-correlation_functions}.
This quantity is defined by

\begin{equation}
\mathcal{A}[\sigma;r]\equiv\frac{1}{L^{2}}\int_{0}^{2\pi}\frac{{\rm d}\theta}{2\pi}\,\int{\rm d}^{2}r^{\prime}\,\left(\sigma_{\boldsymbol{r}+\boldsymbol{r}^{\prime}}\sigma_{\boldsymbol{r}^{\prime}}\right)-\overline{\sigma}^{2},\label{eq:azimuthal_sscf_definition}
\end{equation}
for the $\sigma$ field, where $\boldsymbol{r}=r\left(\cos\theta,\sin\theta\right)$
and $\overline{\sigma}$ is the net $\sigma$-magnetic moment per
site. Similar expressions hold for $\mathcal{A}[\tau;r]$ and $\mathcal{A}[\sigma\tau;r]$.
The autocorrelation functions are nearly linear at short distances
in the linear-log plot of Fig. \ref{fig:Isotropic-spin-spin-correlation_functions},
indicating that $\mathcal{A}(r)\sim{\rm e}^{-r/R}$ for small distances
$r$. The vertical dashed lines denote the distances $R$ where the
spin autocorrelation functions equal ${\rm e}^{-1}$ at the lowest
temperature measured, corresponding to the domains shown in Fig. \ref{fig:domain_decomposition_simulated_annealing}.
It is clear that $R_{\sigma\tau}$ is a factor of $2$-$3$ smaller
than $R_{\sigma},R_{\tau}$. Interestingly, as the system is annealed
and the thermal fluctuations are frozen out, the autocorrelation functions
gradually approach the lowest temperature values, indicating the existence
of temperature-independent length scales for the Baxter and magnetic
variables. These results, although typical, are for a single disorder
configuration. In what follows, we quantitatively study the thermodynamics
and relaxational dynamics of the RBFM after averaging over many different
disorder configurations.

\section{Thermodynamics of the random Baxter field Ashkin-Teller model\label{sec:Thermodynamics-of-the-RBFATM}}

\subsection{Simulation and observables}

We performed Replica-Exchange Wang Landau (WL) Monte Carlo simulations
to obtain the thermodynamics of the RBFM defined in Eq. (\ref{eq:RBFATM_Hamiltonian}).
This method of simulation, as opposed to the Metropolis, heat-bath,
or cluster algorithms, circumvents slowing down near critical points
and at low temperatures. Both the Metropolis and heat-bath algorithms
suffer from critical (supercritical) slowing down near second-order
(first-order) transitions due to the diverging correlation lengths
(phase coexistence). Cluster algorithms, on the other hand, are not
suitable for problems with random-field disorder. The WL algorithm is a
temperature-independent approach that computes the microcanonical
density of states \citep{wang2001efficient_1stpaper,wang2001determining}.
Thermodynamic quantities can be obtained directly from the density
of states. As a result, because temperature is not involved, the WL
algorithm is free from critical or supercritical slowing down.

We employ the massively-parallel Replica-Exchange WL algorithm to
help the simulations converge in the presence of a rugged energy landscape
\citep{vogel2013REWLgeneric,vogel2014REWLscalable,vogel2018REWLtutorial}.
The details of the algorithm are given in Appendix \ref{sec:Numerical-Methods}.
After the thermodynamic properties from each disorder configuration
are obtained, we average over all configurations to obtain the mean
value and standard errors for each observable.

Our analysis focuses on the specific heat and uniform susceptibilities.
We are primarily interested in measurements at a fixed system size
of $L=40$ and two different Baxter couplings: a finite Baxter value
of $K=0.5\,J$ and $K=0$ (Ising-criticality point). We vary the random
Baxter field  strength $\varepsilon$ in Eq. (\ref{eq:random_field_distribution})
to study how the two different break-up length scales affect the disorder-averaged
thermodynamics. All energy values are given in units of $J$ and our
results, when plotted as functions of temperature, are shown within
one standard error represented by a shaded region about the mean.
As thermodynamic quantities can be obtained rapidly from the simulated density of states
for arbitrary temperatures within the WL procedure, we can achieve a high temperature
resolution. In most graphs, the temperature axis has a resolution of $0.001\,J$. When
appropriate, the clean-system transition temperature calculated from Eq.
(\ref{eq:clean_phase_transition_temperature}) will be shown as well
and denoted as $T_{c}^{(0)}$.

The specific heat for a single disorder configuration is calculated
as
\begin{equation}
c_{V}=\frac{1}{NT^{2}}\left(\left\langle E^{2}\right\rangle -\left\langle E\right\rangle ^{2}\right),\label{eq:specific_heat}
\end{equation}
 where $E$ is the energy of the system, $T$ is the temperature,
$N=L^{2}$ is the total number of lattice sites, and $\left\langle \cdot\right\rangle $
represents a canonical thermodynamic average. Denoting the disorder average
(see Appendix \ref{subsec:Disorder-averaging_appendix} for details)
with double brackets $\llbracket \cdot \rrbracket$, the disorder-averaged specific heat is then simply
\begin{equation}
\left\llbracket c_{V}\right\rrbracket =\frac{1}{NT^{2}}\left\llbracket \left\langle E^{2}\right\rangle -\left\langle E\right\rangle ^{2}\right\rrbracket .\label{eq:disorder_avg_specific_heat}
\end{equation}
The number of disorder configurations averaged for each case ranges from  64 to 246;
details are shown
in Tables \ref{tab:Disorder-configurations-averaged} and \ref{tab:Disorder-configurations-averaged-increasing-size}
in Appendix \ref{subsec:Disorder-averaging_appendix}.

There are four susceptibilities of interest in the problem: one for
each magnetic variable, $\sigma$ and $\tau$, one for the composite
Baxter variable $\phi$ representing the nematicity, and one for the
quadrature $\zeta$ of the two Ising variables. We define each of
these quantities as
\begin{align}
\sigma & \equiv\frac{1}{N}\sum_{i=1}^{N}\sigma_{i},\label{eq:sigma_definition}\\
\tau & \equiv\frac{1}{N}\sum_{i=1}^{N}\tau_{i,}\label{eq:tau_definition}\\
\phi & \equiv\frac{1}{N}\sum_{i=1}^{N}\sigma_{i}\tau_{i},\label{eq:baxter_definition}\\
\zeta & \equiv\sqrt{\sigma^{2}+\tau^{2}}.\label{eq:phi_definition}
\end{align}
When there is no risk of confusion, we refer to the Baxter variable
as $\sigma\tau$. Each of these observables has a uniform susceptibility
given by
\begin{equation}
\chi_{\alpha}\equiv\frac{N}{T}\left(\left\langle \alpha^{2}\right\rangle -\left\langle \left|\alpha\right|\right\rangle ^{2}\right),\label{eq:susceptibility_definition}
\end{equation}
for the observable $\alpha$ taken for a single disorder configuration.
Disorder averaging follows the same approach as that for the specific
heat.

\subsection{Finite Baxter coupling}

\begin{figure}
\includegraphics[width=1\columnwidth]{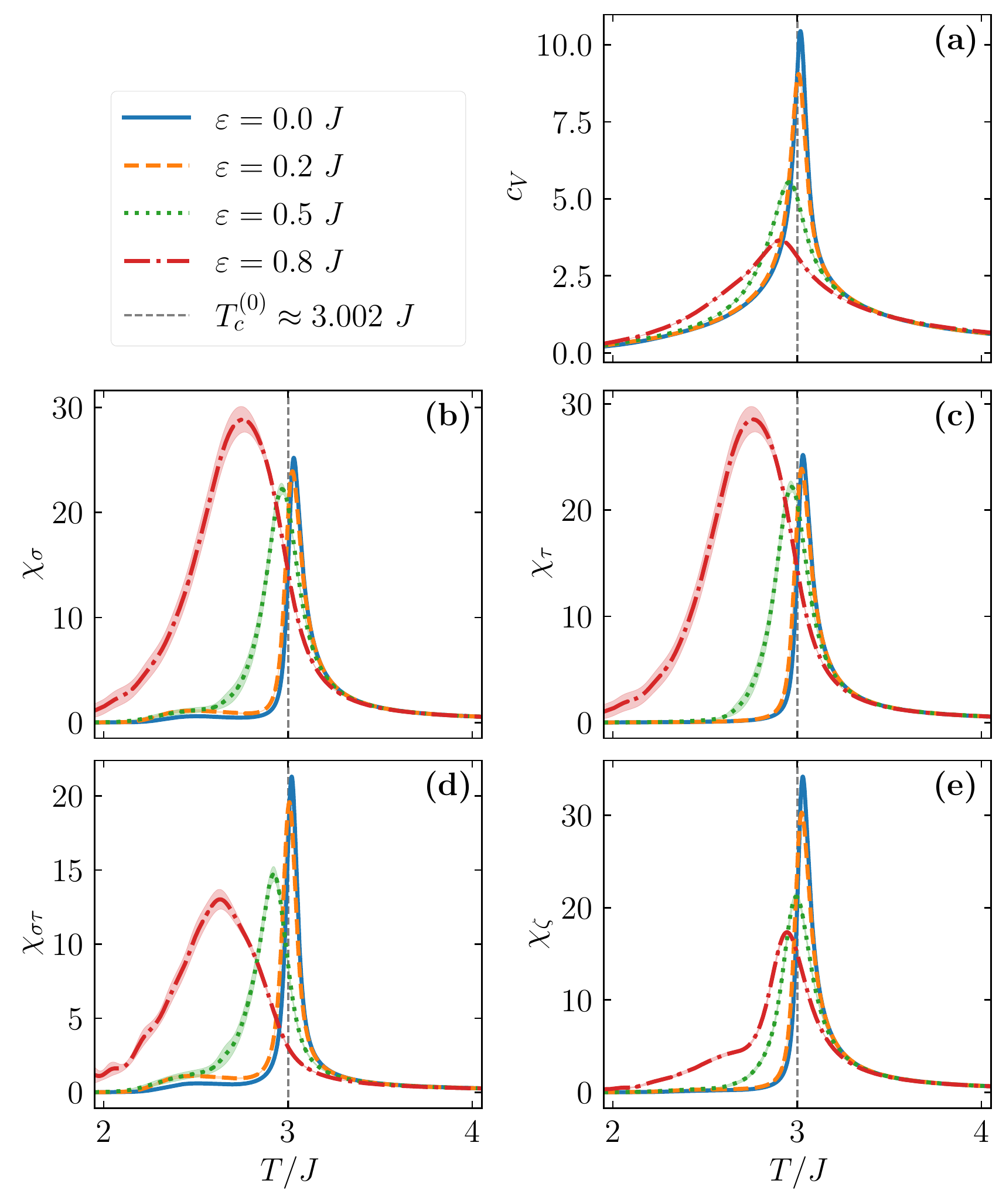}

\caption{Temperature dependence of \textbf{(a)} the specific heat $c_{V}$
and \textbf{(b-e) }the uniform susceptibilities of the RBFM for $K=0.5\,J$
and linear size $L=40$ at different disorder strengths $\varepsilon$.
The susceptibilities are labeled as follows: \textbf{(b)} $\sigma$
susceptibility $\chi_{\sigma}$, \textbf{(c) }$\tau$ susceptibility
$\chi_{\tau}$, \textbf{(d)} the Baxter susceptibility $\chi_{\sigma\tau}$,
and \textbf{(e)} the quadrature susceptibility $\chi_{\zeta}$. \label{fig:K-0.5_cV_and_susceptibilities}}
\end{figure}

\begin{figure}
\includegraphics[width=1\columnwidth]{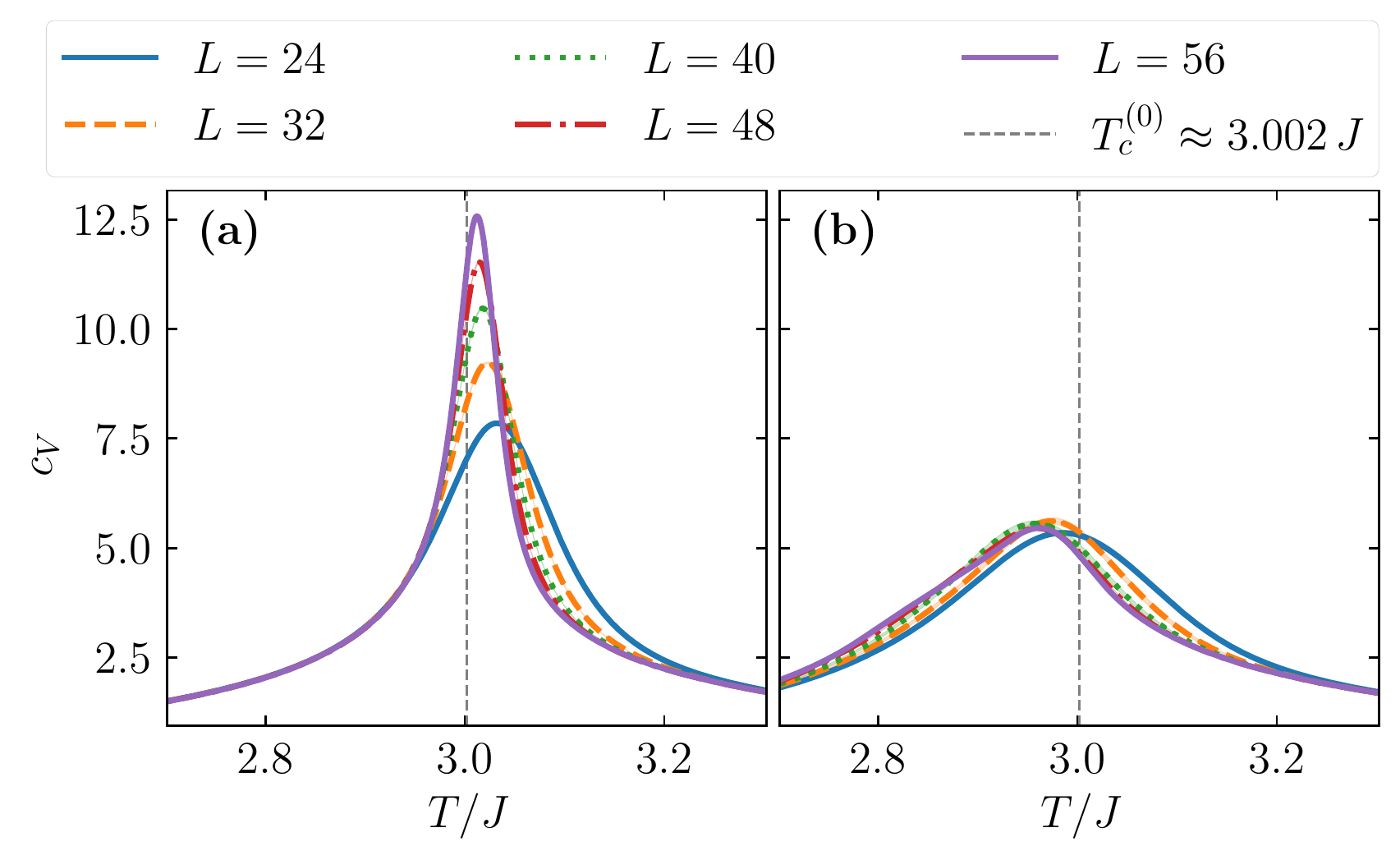}

\caption{Comparison of the specific heat $c_{V}$ of \textbf{(a)} the clean
system $\left(\varepsilon=0\right)$ and \textbf{(b)} the disordered
system $\left(\varepsilon=0.5\,J\right)$ with increasing system size
$L$. The Baxter coupling for both cases was set to $K=0.5\,J$. \label{fig:c_V-variable_system_size}}
\end{figure}

\begin{figure}
\includegraphics[width=1\columnwidth]{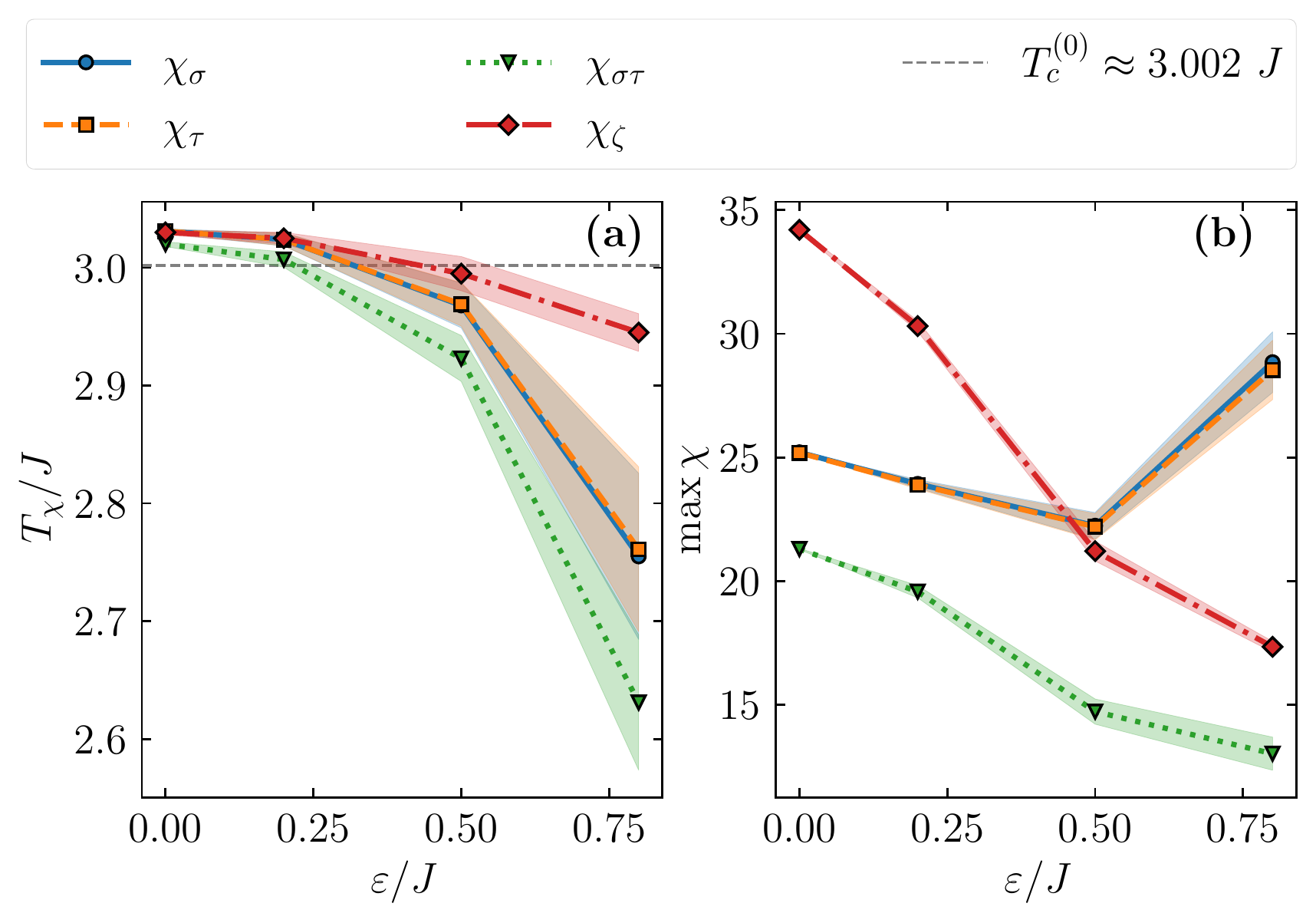}

\caption{\textbf{(a)} Peak susceptibility temperature, $T_{\chi}$, and \textbf{(b)}
peak susceptibility, $\max\chi$, as a function of disorder strength
$\varepsilon$ for $K=0.5\,J$ and $L=40$. The shaded regions represent
$T_{\chi}$ and ${\rm max}\chi$ to within a single error bar. The
lines are guides to the eye. \label{fig:K-0.5_pseudo-Tc_and_peak_value}}
\end{figure}

\begin{figure*}
\noindent\begin{minipage}[t]{1\columnwidth}%
\includegraphics[width=1\columnwidth]{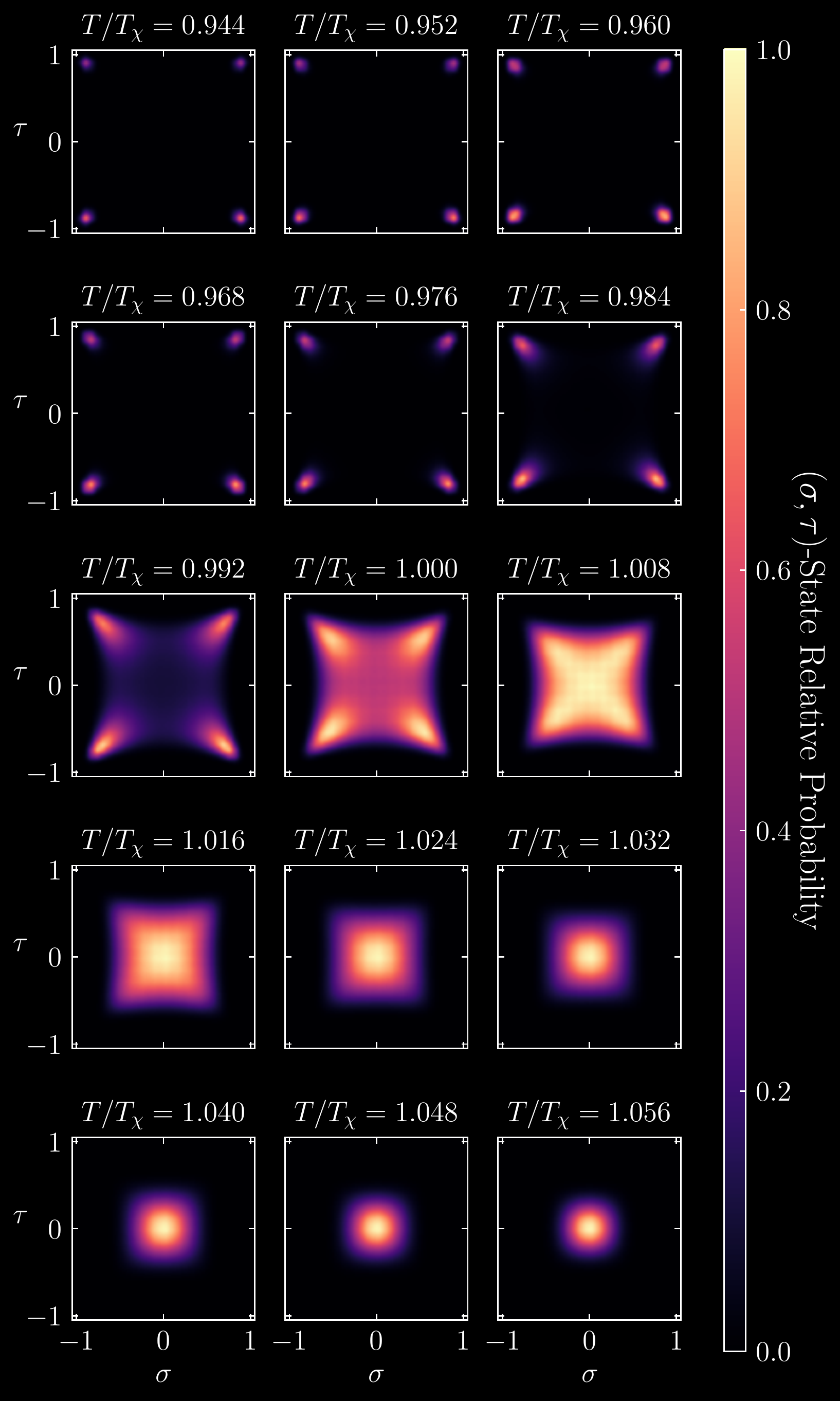}

\caption{Relative thermodynamic probabilities of $\left(\sigma,\tau\right)$-states
for the clean case with $K=0.5\,J$. The central temperature corresponds
to the peak $\chi_{\sigma,\tau}$ position, $T_{\chi}=3.031\,J$.
The highest temperature is $T=3.2\,J$. \label{fig:K-0.5_clean_case_probabilities}}
\end{minipage}\hfill{}%
\noindent\begin{minipage}[t]{1\columnwidth}%
\includegraphics[width=1\columnwidth]{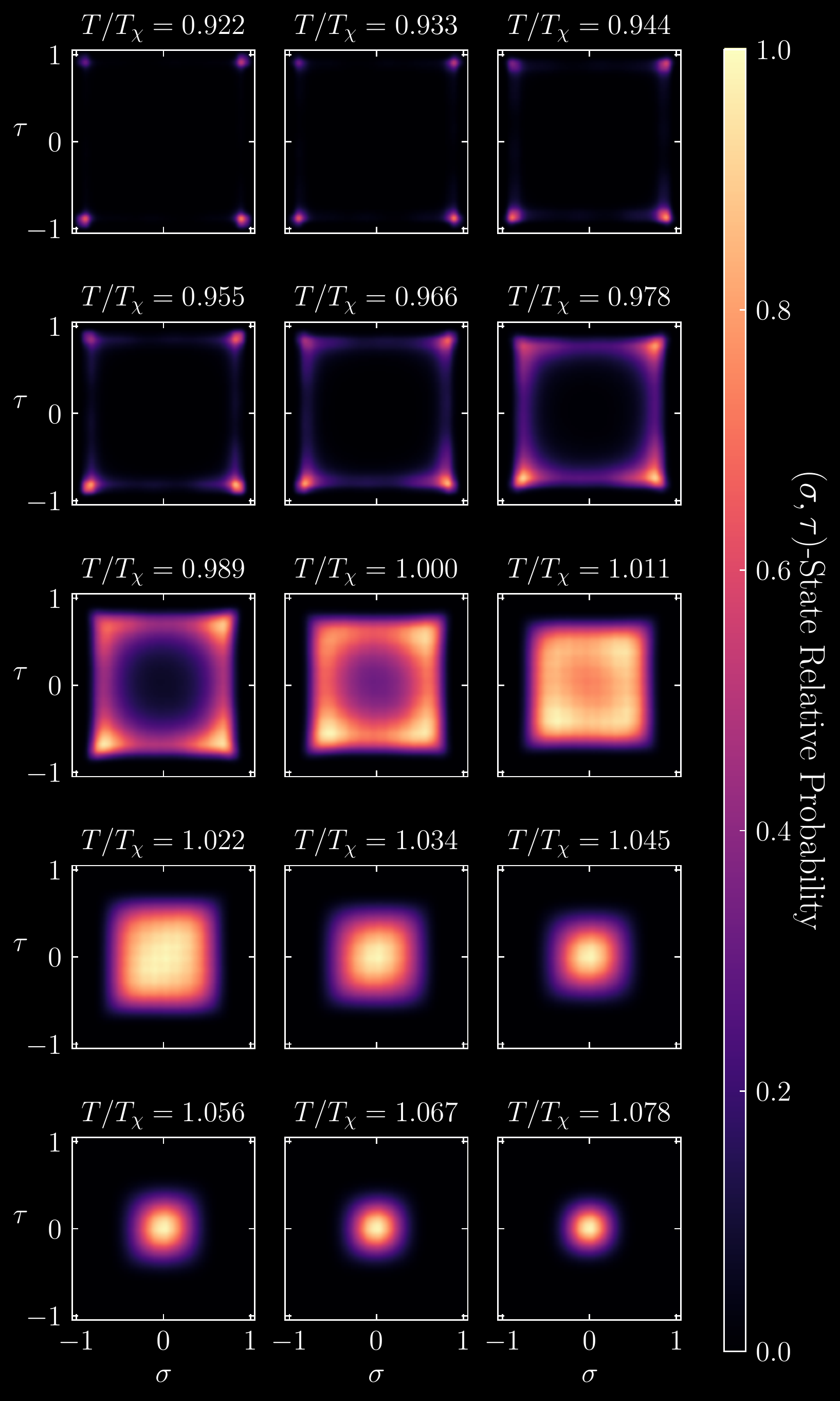}\caption{Relative thermodynamic probabilities of $\left(\sigma,\tau\right)$-states
with $K=0.5\,J$ and a random Baxter field strength of $\varepsilon=0.5\,J$.
The central temperature corresponds to the peak $\chi_{\sigma,\tau}$
position, $T_{\chi}=2.968\,J$. The highest temperature is $T=3.2\,J$,
which is the same as Fig. \ref{fig:K-0.5_clean_case_probabilities}.
\label{fig:K-0.5_h-0.5_probabilities}}
\end{minipage}
\end{figure*}

The specific heat curves for an $L=40$ and $K=0.5\,J$ system and
for different values of the disorder strength $\varepsilon$ are shown
in Fig. \ref{fig:K-0.5_cV_and_susceptibilities}(a). The clean case
$\left(\varepsilon=0\right)$ is shown for comparison. Increasing
the disorder strength leads to a suppression of the specific heat
and to a shift in the peak position towards lower temperatures. These
effects are due to the domain break-up length scale given by Eq. (\ref{eq:breakup_length}),
which decreases with increasing disorder.

This behavior is confirmed by Fig. \ref{fig:c_V-variable_system_size},
which contrasts the specific heat behavior between a clean and a disordered
system with $\varepsilon=0.5\,J$ as the system size $L$ increases.
The peak for the clean case increases and sharpens monotonically, with its position
approaching $T_{c}^{(0)}$. Meanwhile, the peak in the disordered
case saturates and shifts towards $T=0$. This provides further evidence
that the random Baxter field kills the thermodynamic phase transition
of the clean ATM.

Figures \ref{fig:K-0.5_cV_and_susceptibilities}(b-e) show the uniform
susceptibilities for the observables defined in Eqs. (\ref{eq:sigma_definition}-\ref{eq:phi_definition}).
Like the specific heat, the Baxter susceptibility $\chi_{\sigma\tau}$
broadens and its peak is suppressed. A similar behavior is observed
for the quadrature susceptibility $\chi_{\zeta}$. However, the peak
in the two magnetic susceptibilities $\chi_{\sigma}$ and $\chi_{\tau}$
are non-monotonic functions of disorder. They actually increase for
sufficiently large disorder strength, indicating an increase in the
magnetic fluctuations due to the random Baxter field. The behavior
of each susceptibility's peak value ($\max\chi$) and peak temperature
($T_{\chi}$) are shown in Fig. \ref{fig:K-0.5_pseudo-Tc_and_peak_value}
as a function of the disorder strength. We propose that the enhancement
of $\max\chi$ starts when the disorder strength reaches values for which the break-up
length $\ell_{b}$ becomes comparable with or smaller than the system size $L$, such
that the random Baxter field effects become more pronounced. We will
come back to this point in the next section.

In Figs. \ref{fig:K-0.5_clean_case_probabilities} and \ref{fig:K-0.5_h-0.5_probabilities},
we probe the $\left(\sigma,\tau\right)$ configurational space by
showing the joint distribution of the two spin quantities $\sigma$ and $\tau$
for both the clean case and the case with strong disorder $\left(\varepsilon=0.5\,J\right)$,
respectively. These plots thus give the density of $\sigma$ and $\tau$
fluctuations. The highest temperature in each figure (bottom right
panel) was chosen to be the temperature where the magnetic susceptibilities
in Fig. \ref{fig:K-0.5_cV_and_susceptibilities} start to separate
$\left(T\approx3.2\,J\right)$. The temperatures are measured in units
of the peak $\chi_{\sigma,\tau}$ temperature, which we denote by
$T_{\chi}$.

By comparing these figures, one sees that the distributions are nearly
identical at high temperatures, consisting of broad peaks centered
at the origin. However, upon cooling towards $T_{\chi}$, the clean
and disordered distributions behave very differently. In the clean
case, the correlation between the $\sigma$ and $\tau$ variables
enforced by the Baxter exchange $K$ results in peaks along the diagonals
in the $\left(\sigma,\tau\right)$-plane, which become sharper as
the temperature is lowered. Meanwhile, in the disordered case, a nearly
uniform box distribution appears, as shown in the $T=1.022\,T_{\chi}$
panel. This indicates that the $\sigma$ and $\tau$ variables are
nearly uncorrelated, since their joint distribution is a simple product
of two uniform distributions in $\sigma$ and $\tau$. This can be
interpreted as if the Baxter exchange is weakened by the random Baxter
field, enlarging the symmetry of the magnetic fluctuations relative
to the clean system. Moreover, it is consistent with the disorder-promoted
enhancement of $\chi_{\sigma}$ and $\chi_{\tau}$ in Fig. \ref{fig:K-0.5_cV_and_susceptibilities}.

Upon lowering the temperature below $T_{\chi}$, we see that in the
clean case (Fig. \ref{fig:K-0.5_clean_case_probabilities}), the magnetic
variables cluster in four peaks corresponding to the four ground states
of the ATM. In contrast, when the disorder is present (Fig. \ref{fig:K-0.5_h-0.5_probabilities}),
a square-like distribution emerges with empty states in its interior,
as shown in the $T=0.978T_{\chi}$ panel. Such a joint distribution
cannot be described as the product of two independent distributions,
like it could for the case of the uniform square. Instead, they correspond
to uniform fluctuations of one of the magnetic variables while the
other acquires a constant finite value. Therefore, the random Baxter
field induces new correlations between $\sigma$ and $\tau$. The
conclusion we draw from these thermodynamic probability distributions
is that disorder increases the fluctuations of the magnetic degrees
of freedom while correlating them in a distinct way as compared to
the clean case.

\subsection{Zero Baxter coupling}

\begin{figure}
\includegraphics[width=1\columnwidth]{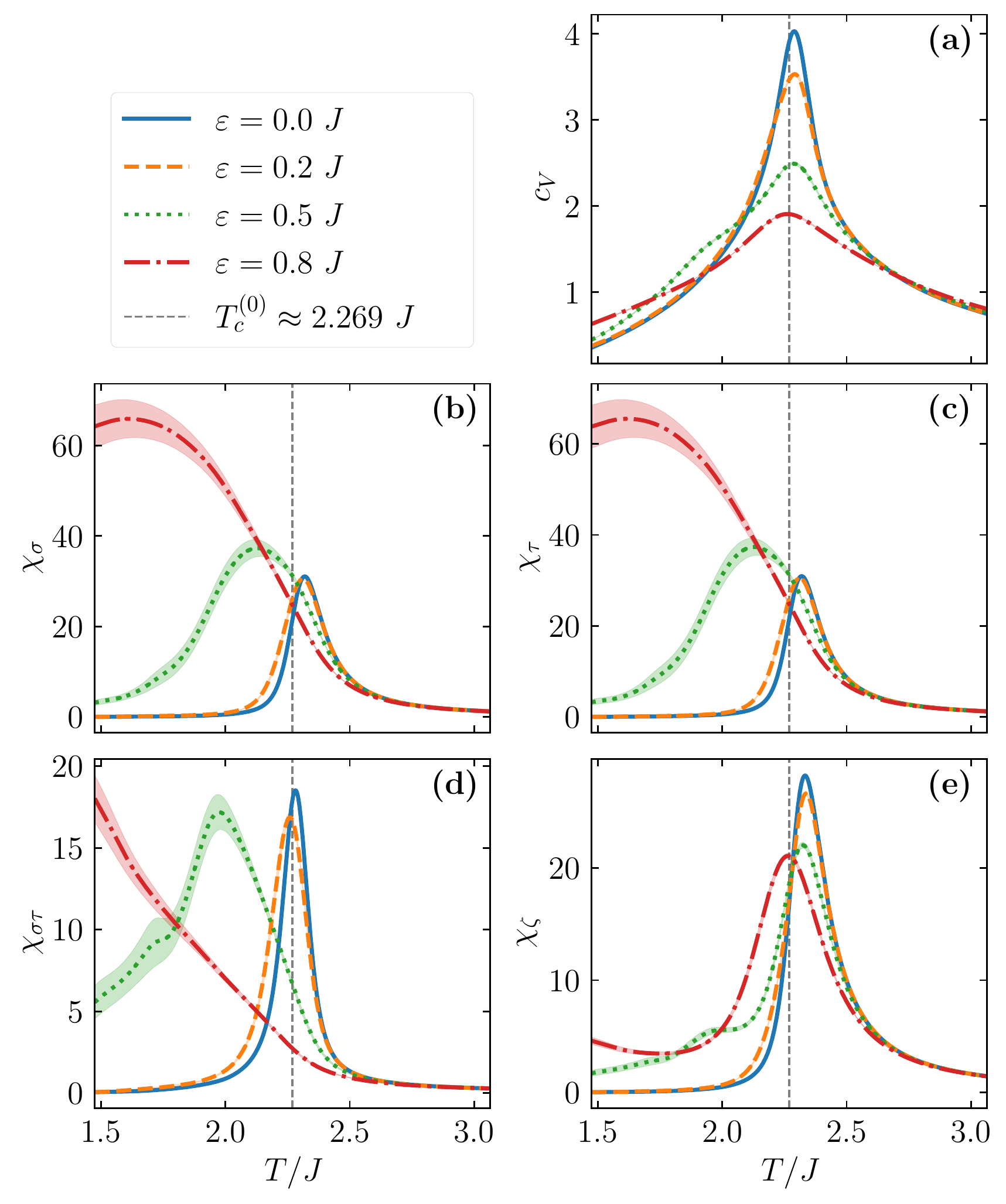}

\caption{Temperature-dependence of \textbf{(a)} the specific heat $c_{V}$
and \textbf{(b-e) }the uniform susceptibilities $(\chi)$ of the RBFM
for $K=0$ and linear size $L=40$ at different disorder strengths
$\varepsilon$. The susceptibilities are labeled as follows: \textbf{(b)}
$\sigma$ susceptibility $\chi_{\sigma}$, \textbf{(c) }$\tau$ susceptibility
$\chi_{\tau}$, \textbf{(d)} the Baxter susceptibility $\chi_{\sigma\tau}$,
and \textbf{(e)} the quadrature susceptibility $\chi_{\zeta}$. \label{fig:K-0.0_cV_and_susceptibilities}}
\end{figure}

\begin{figure}
\includegraphics[width=1\columnwidth]{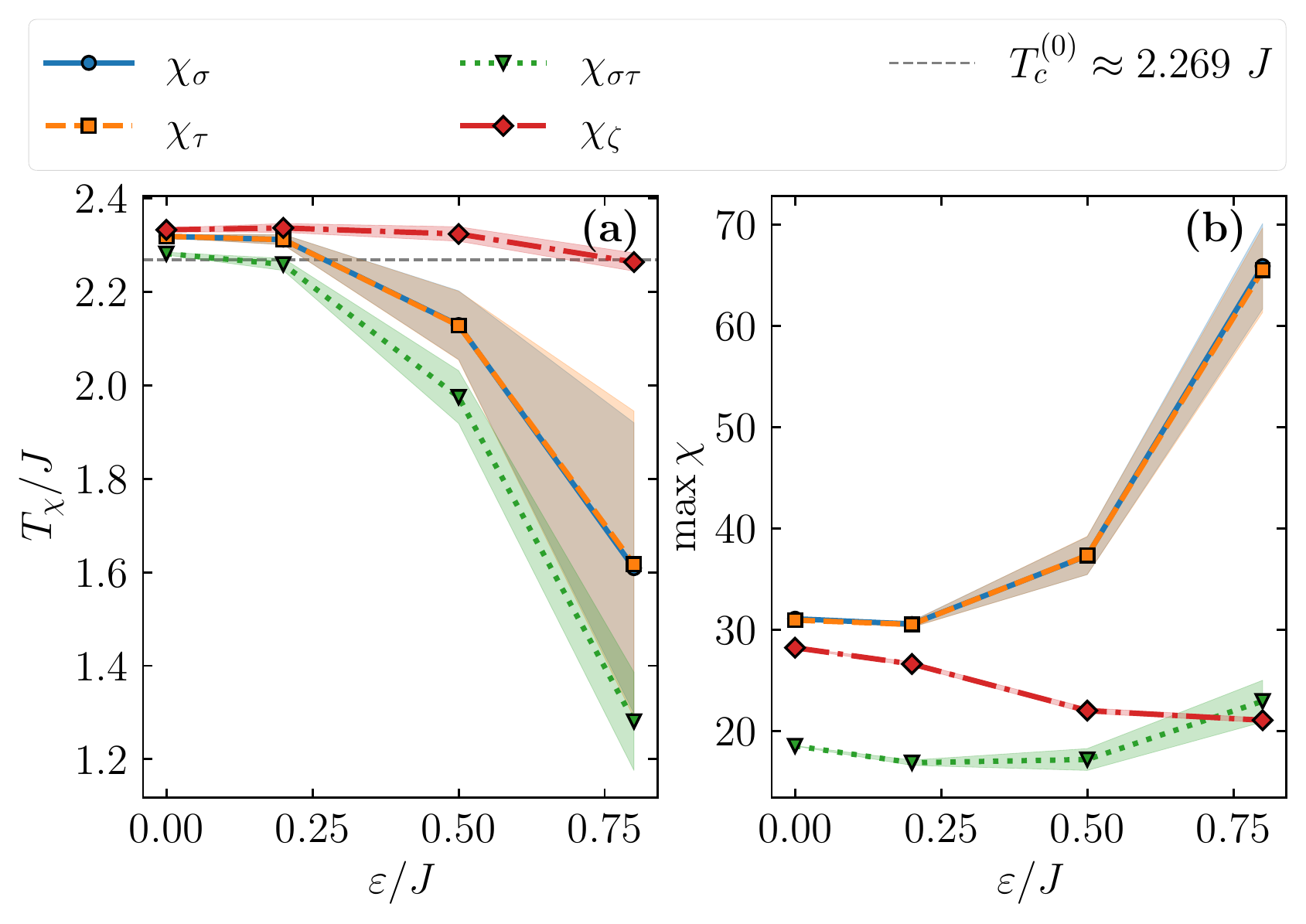}

\caption{\textbf{(a)} Peak susceptibility temperature, $T_{\chi}$, and \textbf{(b)}
peak susceptibility, $\max\chi$, as a function of disorder strength
$\varepsilon$ for $K=0$ and $L=40$. The shaded regions represent
$T_{\chi}$ and ${\rm max}\chi$ to within a single error bar. The
lines are guides to the eye. \label{fig:K-0.0_pseudo-Tc_and_peak_value}}
\end{figure}

\begin{figure*}
\noindent\begin{minipage}[t]{1\columnwidth}%
\includegraphics[width=1\columnwidth]{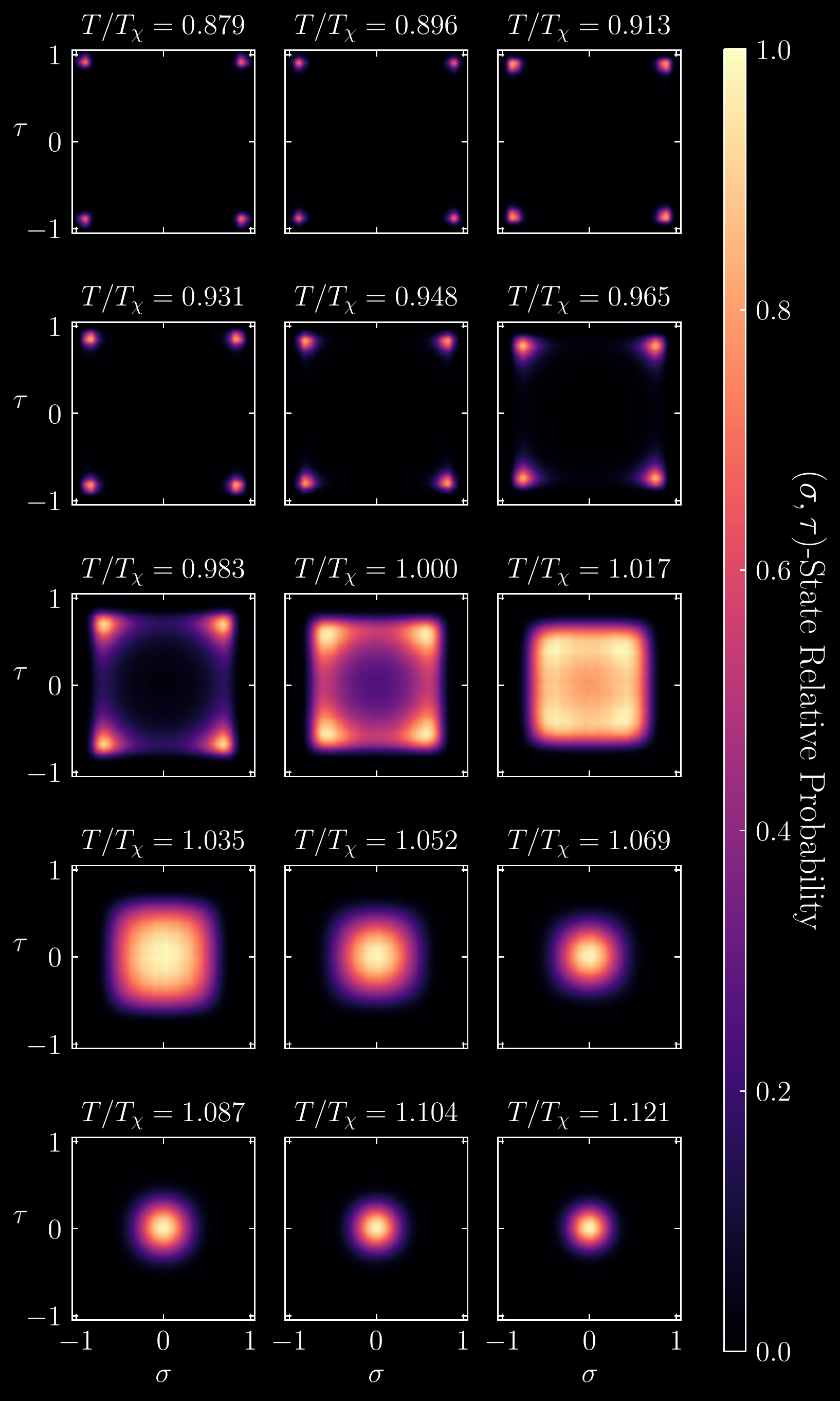}

\caption{Relative thermodynamic probabilities of $\left(\sigma,\tau\right)$-states
for the clean case for $K=0$. The central temperature corresponds
to the peak $\chi_{\sigma,\tau}$ position, $T_{\chi}=2.319\,J$.
The highest temperature is $T=2.6\,J$. \label{fig:K-0.0_clean_case_probabilities}}
\end{minipage}\hfill{}%
\noindent\begin{minipage}[t]{1\columnwidth}%
\includegraphics[width=1\columnwidth]{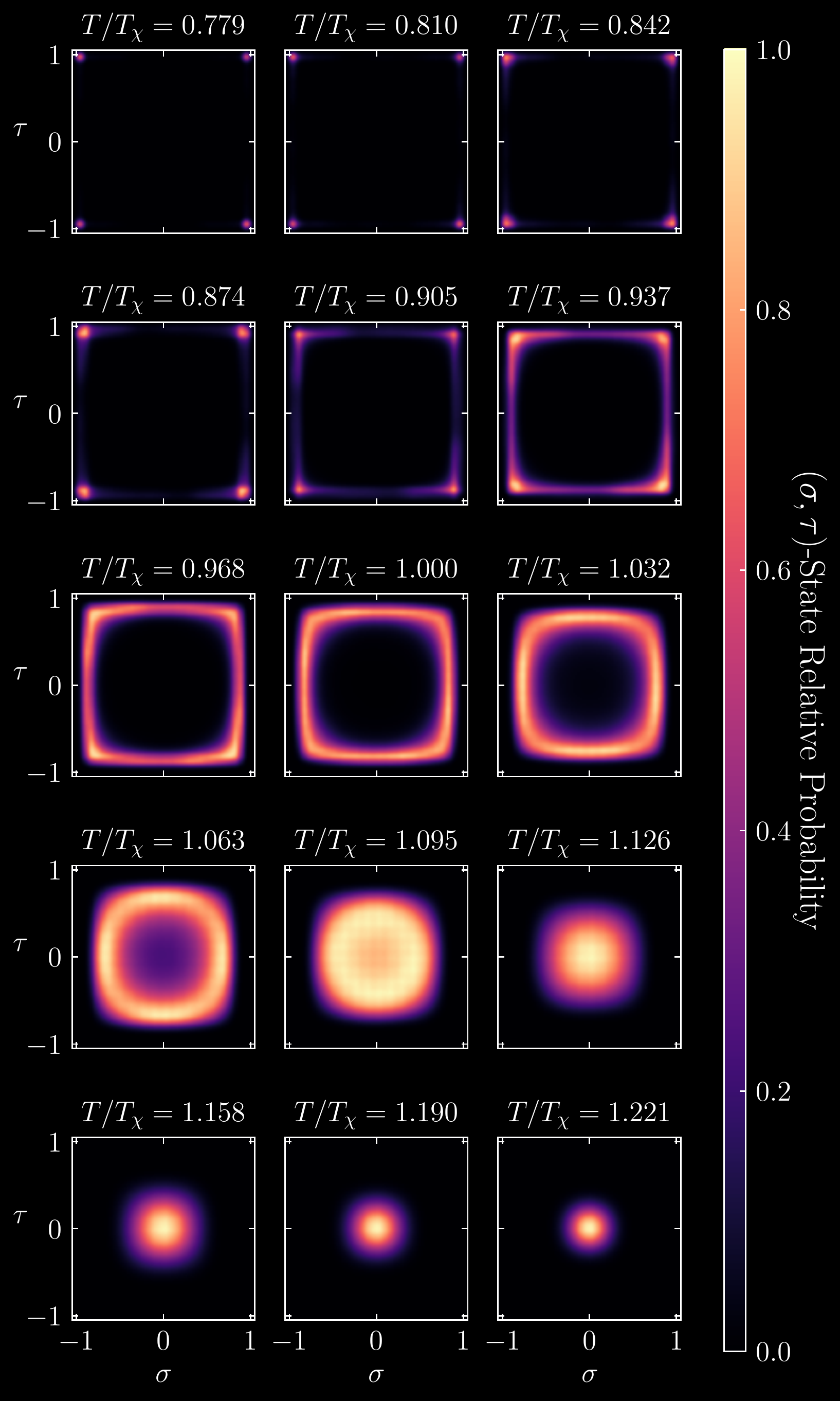}

\caption{Relative thermodynamic probabilities of $\left(\sigma,\tau\right)$-states
for $K=0$ with $\varepsilon=0.5\,J$. The central temperature corresponds
to the peak $\chi_{\sigma,\tau}$ position, $T_{\chi}=2.129\,J$.
The highest temperature is $T=2.6\,J$ as it is in Fig. \ref{fig:K-0.0_clean_case_probabilities}.\label{fig:K-0.0_h-0.5_probabilities}}
\end{minipage}
\end{figure*}

Our numerical results for systems of size $L=40$ in Fig. \ref{fig:K-0.5_pseudo-Tc_and_peak_value}(b)
show that when the disorder strength becomes comparable to the Baxter
exchange scale, $K/J$, the magnetic fluctuations grow. To shed light
on this result, and to disentangle the correlations between $\sigma$
and $\tau$ induced by disorder (shown in Fig. \ref{fig:K-0.5_h-0.5_probabilities})
from the correlations enforced by $K$ (shown in Fig. \ref{fig:K-0.5_clean_case_probabilities}),
we simulate the $K=0$ RBFM. It corresponds to two clean Ising models
coupled to each other only locally by the random Baxter field.

The specific heat for the RBFM with linear size $L=40$ and $K=0$
is shown in Fig. \ref{fig:K-0.0_cV_and_susceptibilities}(a). While
the specific heat peak is suppressed with increasing disorder, it
stays essentially at the clean transition temperature $T_{c}^{(0)}$
of the 2D Ising model. This contrasts with the $K=0.5\,J$ case (Fig.
\ref{fig:K-0.5_cV_and_susceptibilities}(a)), where the peak position
shifts to lower temperatures.

On the other hand, the uniform susceptibilities in Fig. \ref{fig:K-0.0_cV_and_susceptibilities}(b-e)
for $K=0$ show a qualitatively similar behavior with increasing disorder
as they did for $K=0.5\,J$. Quantitatively, however, the magnetic
fluctuations overcome the clean-system susceptibility for smaller
$\varepsilon$ values, as shown in the plots of the susceptibility's peak
value ($\max\chi$) and peak temperature ($T_{\chi}$) in Fig. \ref{fig:K-0.0_pseudo-Tc_and_peak_value}.
This is consistent with the break-up length $\ell_{b}$ in Eq. (\ref{eq:breakup_length})
being suppressed in the $K=0$ case, thus reaching the system's size
$L$ for smaller disorder strength values than in the $K=0.5J$ case.\textcolor{red}{{}
}We also note that, unlike the $K=0.5\,J$ case, the Baxter fluctuations
eventually exceed the clean system's fluctuations at large enough
disorder values.

Fig. \ref{fig:K-0.0_clean_case_probabilities} shows the relative
thermodynamic probability of a $\left(\sigma,\tau\right)$-state for
$K=0$ and $\varepsilon=0$, while the disordered case with $\varepsilon=0.5\,J$
is shown in Fig. \ref{fig:K-0.0_h-0.5_probabilities}. The distribution
at peak temperature $T/T_{\chi}=1$ for the clean system in Fig. \ref{fig:K-0.0_clean_case_probabilities}
is qualitatively similar to that in Fig. \ref{fig:K-0.5_h-0.5_probabilities}
for the disordered system with $K=\varepsilon=0.5\,J$. Indeed, both
cases show a lack of statistical weight near the paramagnetic state
$\left(\sigma,\tau\right)=\left(0,0\right)$. Moreover, at temperatures
just above $T_{\chi}$, both distributions have a nearly uniform-box
structure, indicating that $\sigma$ and $\tau$ are nearly uncorrelated.
These comparisons support the notion that the random Baxter field
effectively renormalizes the finite Baxter coupling down.

This is not the only effect of the random Baxter field. Comparing
Fig. \ref{fig:K-0.0_h-0.5_probabilities}, which refers to the disordered
system $\left(K=0,\,\varepsilon=0.5\,J\right)$, with Fig. \ref{fig:K-0.5_h-0.5_probabilities},
which refers to the disordered system with finite Baxter coupling
$\left(K=0.5\,J,\,\varepsilon=0.5\,J\right)$, the emergence of disorder-induced
correlations between $\sigma$ and $\tau$ near $T_{\chi}$ is much
more transparent in the former. Specifically, as the system cools
down from the high temperature paramagnetic phase, the statistical
weight moves radially away from the paramagnetic state and eventually
forms a hollow square-like shape, which is sharper than the similar
hollow-square shape in Fig. \ref{fig:K-0.5_h-0.5_probabilities} for
$K=0.5\,J$. Once again, this square-like joint-distribution cannot
be factorized into two separate distributions of $\sigma$ and $\tau$,
and it is absent completely in the clean case with $K=0$. This means
that the random Baxter field introduces correlations between the magnetic
variables. Along each side of the square-like distribution, one of
the magnetic variables is fixed at some value, while the other thermally
fluctuates. This behavior satisfies the random Baxter field constraint
of $\left\langle \sigma\tau\right\rangle =0$ by making, for example,
$\left\langle \sigma\right\rangle =0$ while $\left\langle \tau\right\rangle \neq0$
along the upper and lower sides of the squares.

\subsection{Comparison with the random-field Ising model\label{subsec:Comparison-with-the_RFIM}}

\begin{figure}
\includegraphics[width=1\columnwidth]{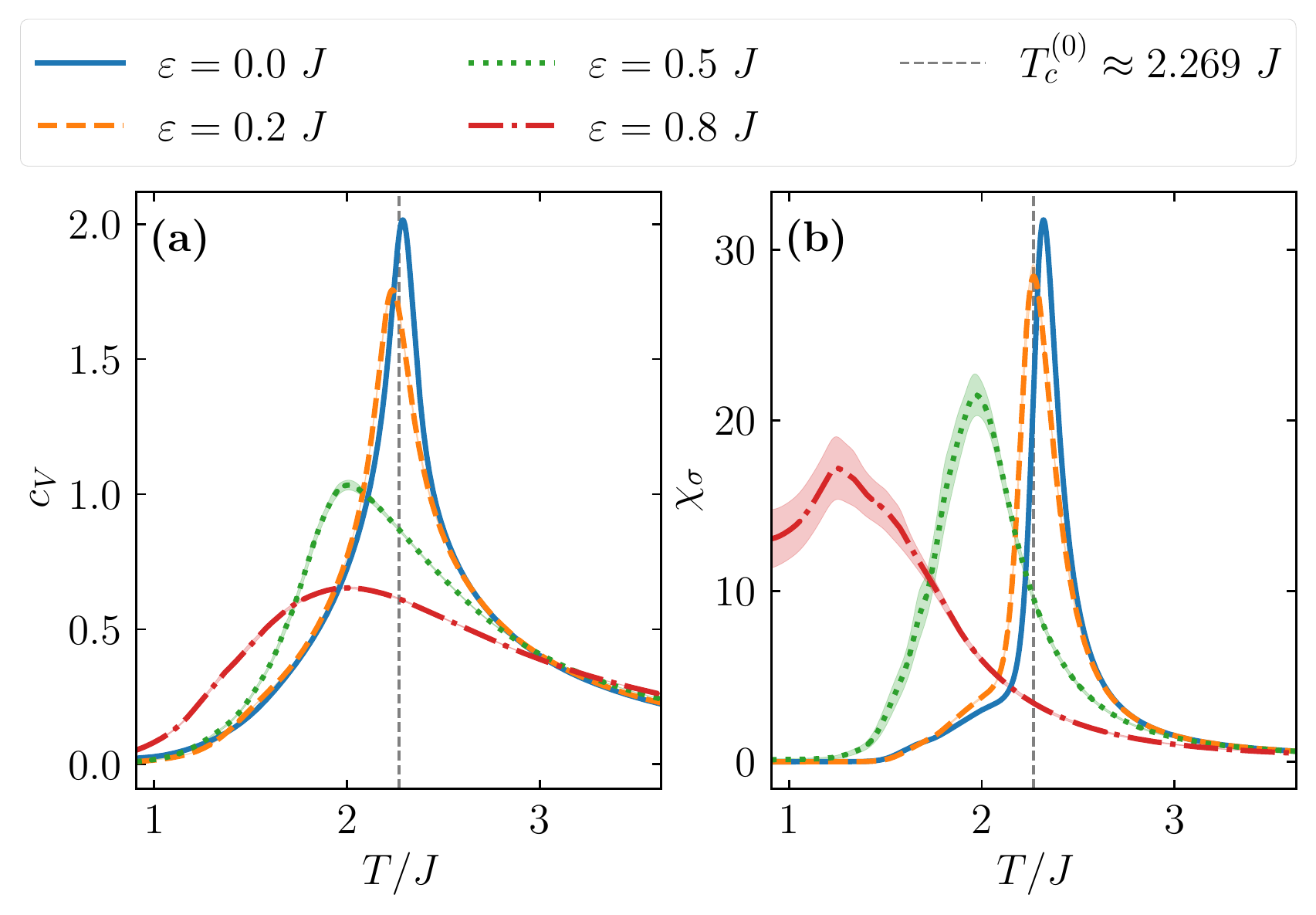}

\caption{Disorder-averaged thermodynamic quantities of the RFIM with linear
size $L=40$, measured as functions of temperature $T$, for various
random field strengths $\varepsilon$. \textbf{(a)} The specific heat
$c_{V}$. \textbf{(b)} The magnetic susceptibility $\chi_{\sigma}$.
\label{fig:RFIM_thermodynamics}}
\end{figure}

As discussed in the Introduction, while the RBFM describes the impact
of random strain on composite nematicity arising from an underlying
magnetic/charge stripe order, the RFIM provides a suitable framework
to describe the case in which nematicity is a primary instability
of the system. It is therefore valuable to compare the properties
of the RBFM with those of the RFIM, particularly when $K=0$, since
in this case the clean version of the RBFM belongs to the Ising universality
class -- similarly to the clean version of the RFIM. On the one hand,
the RBFM displays the hallmark feature of the RFIM: the break-up of
long-range nematic order into nematic domains. On the other hand,
some of the thermodynamic properties of the RBFM differ from those
of the RFIM.

Consider, for instance, the specific heat and the magnetic susceptibility
for an $L=40$ RFIM shown in Fig. \ref{fig:RFIM_thermodynamics},
which we obtained from our Replica-Exchange WL simulation. The peaks
of both $c_{V}$ and $\chi_{\sigma}$ are suppressed and the corresponding
peak temperatures shift towards $T=0$. In the RBFM for $K=0$, as
shown in Fig. \ref{fig:K-0.0_cV_and_susceptibilities}, the only thermodynamic
quantities that share qualitative similarities to those of the RFIM
are the Baxter and quadrature susceptibilities ($\chi_{\sigma\tau}$
and $\chi_{\zeta}$). The magnetic susceptibilities ($\chi_{\sigma}$
and $\chi_{\tau}$) and, more importantly, the specific heat $c_{V}$
are different. Thus, the disorder physics in the RBFM must include
effects with no counterpart in the RFIM.

To further illuminate this issue, we rewrite the RBFM Hamiltonian
in Eq. (\ref{eq:RBFATM_Hamiltonian}) in terms of the local Baxter
Ising variable $\phi_{i}=\sigma_{i}\tau_{i}$:
\begin{align}
H & =-J\sum_{\left\langle ij\right\rangle }\sigma_{i}\sigma_{j}-K\sum_{\left\langle ij\right\rangle }\phi_{i}\phi_{j}-\sum_{i}\varepsilon_{i}\phi_{i}\nonumber \\
 & \quad-J\sum_{\left\langle ij\right\rangle }\phi_{i}\phi_{j}\sigma_{i}\sigma_{j}\label{eq:RBFATM_Rewritten}
\end{align}

Written in this way, the RBFM Hamiltonian takes the form of a clean
Ising model (IM) in $\sigma$ coupled to a RFIM in $\phi$ through
a four-spin operator. It is this four-spin operator that creates the
correlations in the magnetic variables and gives rise to effects beyond
the RFIM. Such an operator is particularly interesting because it
effectively generates a bond-dependent exchange for both the $\sigma$
IM and the $\phi$ RFIM components of the RBFM. The effective $\sigma$
exchange is given by $J_{ij}^{(\sigma)}\equiv-J\left(1+\phi_{i}\phi_{j}\right)$
which is ferromagnetic when the bond connects Baxter variables in
the same domain with $\phi_{i}\phi_{j}=1$. However, when the bond
is across a Baxter domain wall such that $\phi_{i}\phi_{j}=-1$, then
$J_{ij}^{(\sigma)}=0,$ i.e. the magnetic degrees of freedom separated
by this bond are independent. This again illustrates that domain break-up
in the Baxter variable necessarily leads to domain break-up in the
magnetic variables, and shows how the magnetic break-up length is
larger than the Baxter one.

\begin{figure}
\includegraphics[width=1\columnwidth]{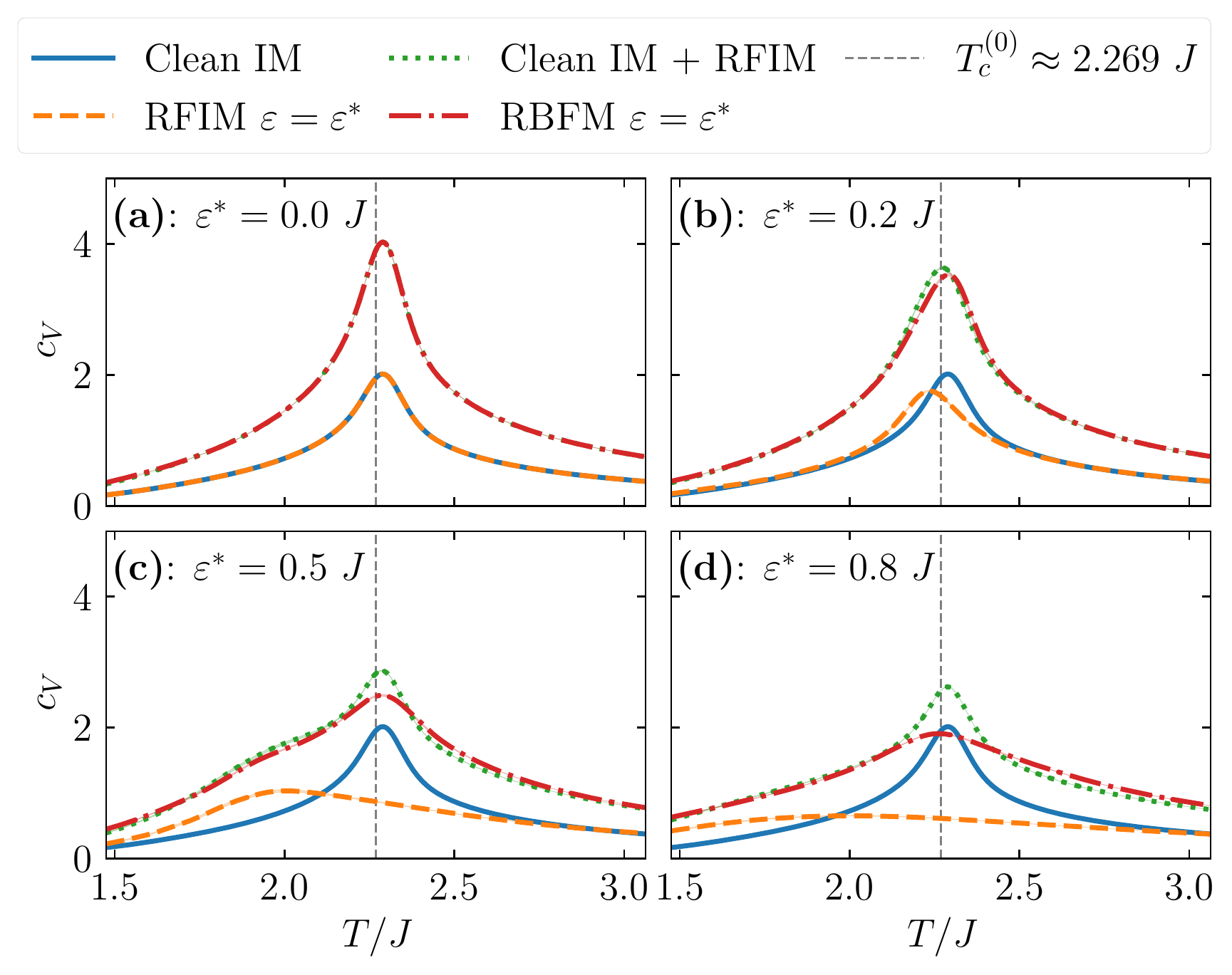}

\caption{Comparison of the RFIM and the RBFM with $L=40$ and $K=0$ over various
disorder strengths. These plots show the specific heat $c_{V}$ as
a function of temperature $T$ for the clean Ising model (Clean IM);
the RFIM at $\varepsilon=\varepsilon^{*}$, where $\varepsilon^{*}$
is a numerical value defined in each figure; the sum of the specific
heats of the Clean IM and the RFIM (Clean IM + RFIM); and the RBFM
at zero Baxter coupling. \textbf{(a)} $\varepsilon^{*}=0.0\,J$, \textbf{(b)}
$\varepsilon^{*}=0.2\,J$, \textbf{(c)} $\varepsilon^{*}=0.5\,J$,
and \textbf{(d)} $\varepsilon^{*}=0.8\,J$. In the clean case \textbf{(a)},
the Ashkin-Teller model is equivalent to two clean Ising models, as
shown by the complete agreement between the green dotted curve and
the red dash-dotted curve. \label{fig:Comparison-of-the_RFIM-RBFATM}}
\end{figure}

Conversely, the effective Baxter exchange is given by $J_{ij}^{(\phi)}\equiv-K-J\sigma_{i}\sigma_{j}$.
Thus, if neighboring magnetic spins belong to the same domain with
$\sigma_{i}\sigma_{j}=1$, then the Ising Baxter degrees of freedom
experience a ferromagnetic exchange. Meanwhile, if the neighboring
magnetic spins are separated by a magnetic domain wall such that $\sigma_{i}\sigma_{j}=-1$,
then the Ising Baxter degrees of freedom experience an antiferromagnetic
exchange. This additional frustration for the Baxter variables leads
to more break-up, providing a feedback mechanism on the composite
degrees of freedom that is lacking in the conventional application
of the RFIM. But these effects only become clear when both the $\phi$
and $\sigma$ break-ups are present.

To test this idea, Fig. \ref{fig:Comparison-of-the_RFIM-RBFATM} shows
the evolution of the specific heat of the RBFM as the disorder strength
is increased, and compares it to that of a clean IM, the RFIM, and
the sum of the clean IM and the RFIM. For small disorder strengths,
there is barely any Baxter domain break-up, which explains the small
difference between the RBFM and the sum of the two Ising models. As
the disorder strength increases, however, and domain break-up in both
variables is more common, the splitting of the RBFM curve from the
curves obtained from the sum of a clean IM and a RFIM becomes clearer.
While some of the behavior of the RBFM specific heat is captured by
the sum, such as the frozen peak temperature of the specific heat
and the bump at low $T$, the suppression of the specific heat peak
is not. This suppression indeed highlights the role that the random
Baxter field has in correlating the primary magnetic degrees of freedom
in a way absent in the RFIM.

\section{Relaxational Dynamics of the random Baxter field Ashkin-Teller model\label{sec:Dynamics-at-zero-Baxter-coupling}}

\subsection{Dynamical simulation details}

Our Replica-Exchange WL results reveal that for strong enough disorder,
when the Baxter coupling is effectively suppressed, the two magnetic
variables remain coupled to each other through the random Baxter field.
As shown in Fig. \ref{fig:K-0.0_h-0.5_probabilities}, these disorder-induced
correlations force only one magnetic variable to fluctuate at a time,
enlarging the magnetic break-up lengths over the nematic one.

The two domain break-up length scales for the magnetic and Baxter
domains should influence their dynamics as well. For example, if one
were to assume relaxational domain dynamics of the Arrhenius law form,
then the typical frequency of switching events would scale as $\exp\left(-\lambda\ell/T\right)$,
where $\lambda$ is the surface tension of a domain wall and $\ell$
is the domain size. Then, when two typical domain sizes exist in the
problem, there should be two typical frequencies, with the faster
arising from the smaller Baxter domains and the slower one, from the
larger magnetic domains. To verify this expectation, we simulate here
random-strain effects on the relaxational dynamics of the RBFM.

Metropolis Monte Carlo \citep{metropolis1953} simulations were run
for the RBFM to simulate the relaxational dynamics of the magnetic
fluctuations for the $K=0$ case. The Metropolis algorithm simulates
the dynamics of thermally-activated excitations of equilibrium systems.
We use the time-records generated by the Metropolis Markov chains
as a proxy of the time-evolution of these excitations. This is justified
if the model dynamics belongs to Hohenberg and Halperin's \emph{Model
A} classification \citep{hoenberg_halperin1977modelA,chaikin_lubensky_1995}.
Considering that the stripe magnetic phase has a non-conserved order
parameter, and that neither the magnetic nor the nematic order parameters
are coupled to conserved quantities, the domain dynamics of the model
is indeed expected to be relaxational and fall within the Model A
classification.

The Monte Carlo timescale was set by a single sweep, defined by $2L^{2}$
single-spin Metropolis updates. A single Monte Carlo sweep then maps
onto the shortest physical real-time scale of the problem. This is
associated with the expected time it takes for a single Ising spin
to flip from one projection to the other. We note that, in iron pnictides,
the typical relaxational timescales observed in the nematic-magnetic
phase is of the order of tens of picoseconds \citep{Patz2014}.

The clean system was simulated at the clean 2D Ising transition temperature
$T=2.269\,J$ while the disordered system's temperature was set to
$T_{\chi}=2.129\,J$ (see Figs. \ref{fig:K-0.0_pseudo-Tc_and_peak_value}
and \ref{fig:K-0.0_h-0.5_probabilities}). These values were chosen
to compare critical magnetic fluctuations in the clean system with
the disorder-enhanced fluctuations of the RBFM.

The time records of $\sigma$, $\tau$, $\sigma\tau$, and $\zeta$
were taken for clean $\left(\varepsilon=0\right)$ and disordered
systems $\left(\varepsilon=0.5\,J\right)$ with $L=80$ and $K=0$.
The disordered systems each had different quenched disorder configurations.
Each Markov chain was thermalized for $2^{18}\approx2.6\times10^{5}$
sweeps and then the time records were measured for $L_{t}=2^{22}\approx4.2\times10^{6}$
sweeps. The power spectral density (PSD) was averaged 100 times for
the clean case and over 200 disorder configurations for the disordered
one. For a given time record $x=x(t)$ with time-average $\overline{x}$,
the PSD is defined by
\begin{equation}
\mathcal{P}[x(t);f]\equiv\frac{1}{L_{t}}\tilde{x}^{*}(f)\tilde{x}(f)-\overline{x}^{2}L_{t}\delta_{f,0},\label{eq:PSD_definition}
\end{equation}
where $f$ has units of frequency and $\tilde{x}(f)$ is the discrete
Fourier transform of $x(t)$. The PSD above is the cosine-transform
of the autocorrelation function of $x(t)$ given by
\begin{equation}
\mathcal{A}[x(t);t]=\frac{1}{L_{t}}\int{\rm d}t^{\prime}\,\left[x\left(t+t^{\prime}\right)x\left(t^{\prime}\right)\right]-\overline{x}^{2}.\label{eq:autocorrelation_function}
\end{equation}
The PSD is therefore the spectrum of the fluctuations contained in
the time record. For fast timescales (large frequencies), the PSDs
typically have power-law behavior. Meanwhile, for timescales longer
than the longest switching time in the time record (small frequencies),
the PSDs typically saturate \citep{MacDonald_D_K_C_2006-06-23,dutta_horn1981noise_review,weissman1988noise_review}.

\begin{figure*}
\noindent\begin{minipage}[t]{1\columnwidth}%
\includegraphics[width=1\columnwidth]{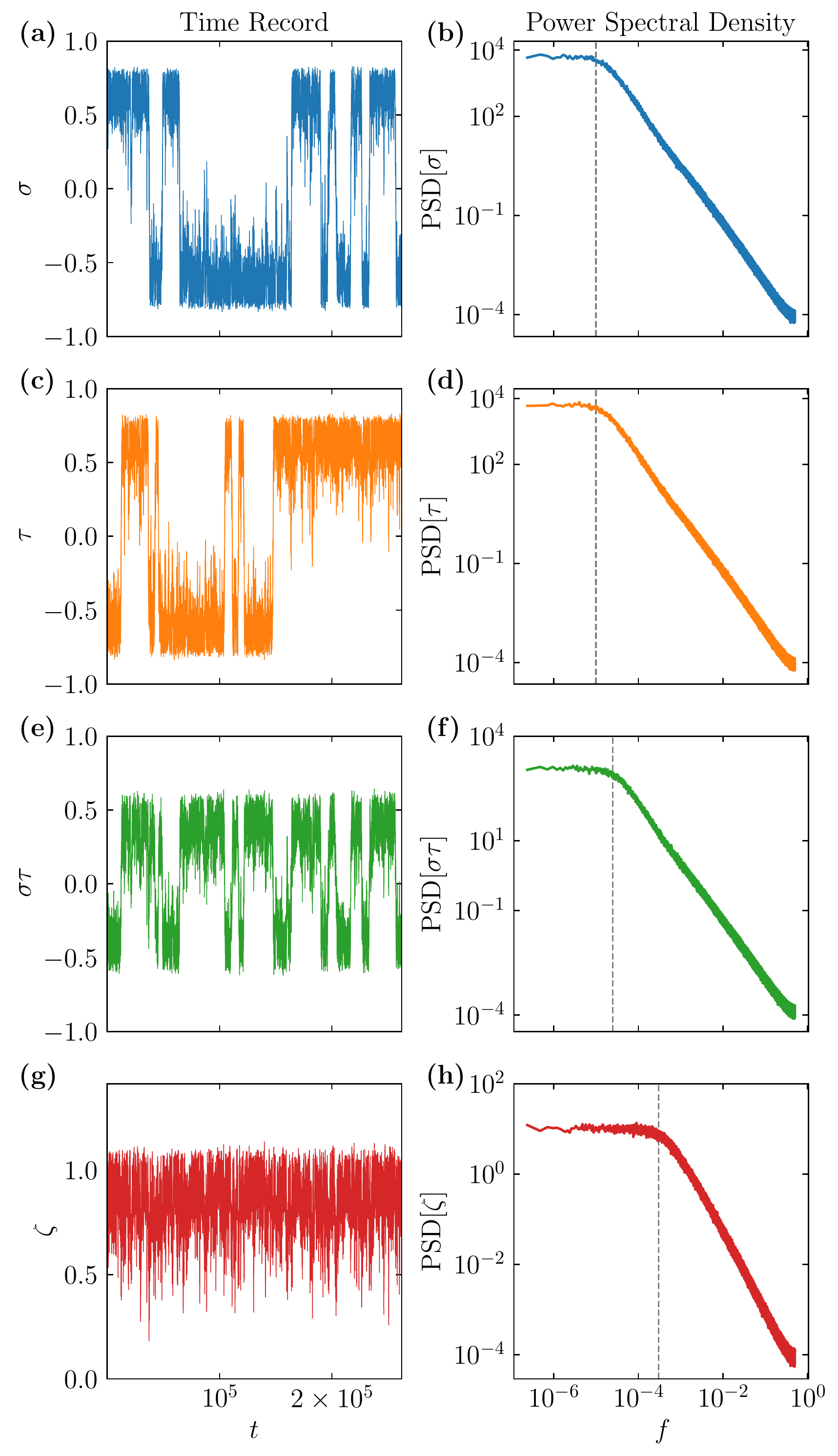}

\caption{Metropolis evolution of the RBFM for a system size $L=80$ and $K=0$
for the clean case $\left(\varepsilon=0\right)$ at the transition
temperature $T_{c}^{(0)}\approx2.269\:J$. We show a typical time
record as a function of Monte Carlo sweeps $t$ and the average power
spectral density (PSD) as a function of frequency $f$ for the \textbf{(a,
b)} $\sigma$, \textbf{(c, d)} $\tau$, \textbf{(e, f)} Baxter $\sigma\tau$,
and \textbf{(g, h)} quadrature $\zeta$ variables. The vertical lines
correspond approximately to the frequencies at which the PSDs crossover
from a plateau to a power-law behavior. \label{fig:Metropolis-evolved-time-records_h-0.0}}
\end{minipage}\hfill{}%
\noindent\begin{minipage}[t]{1\columnwidth}%
\includegraphics[width=1\columnwidth]{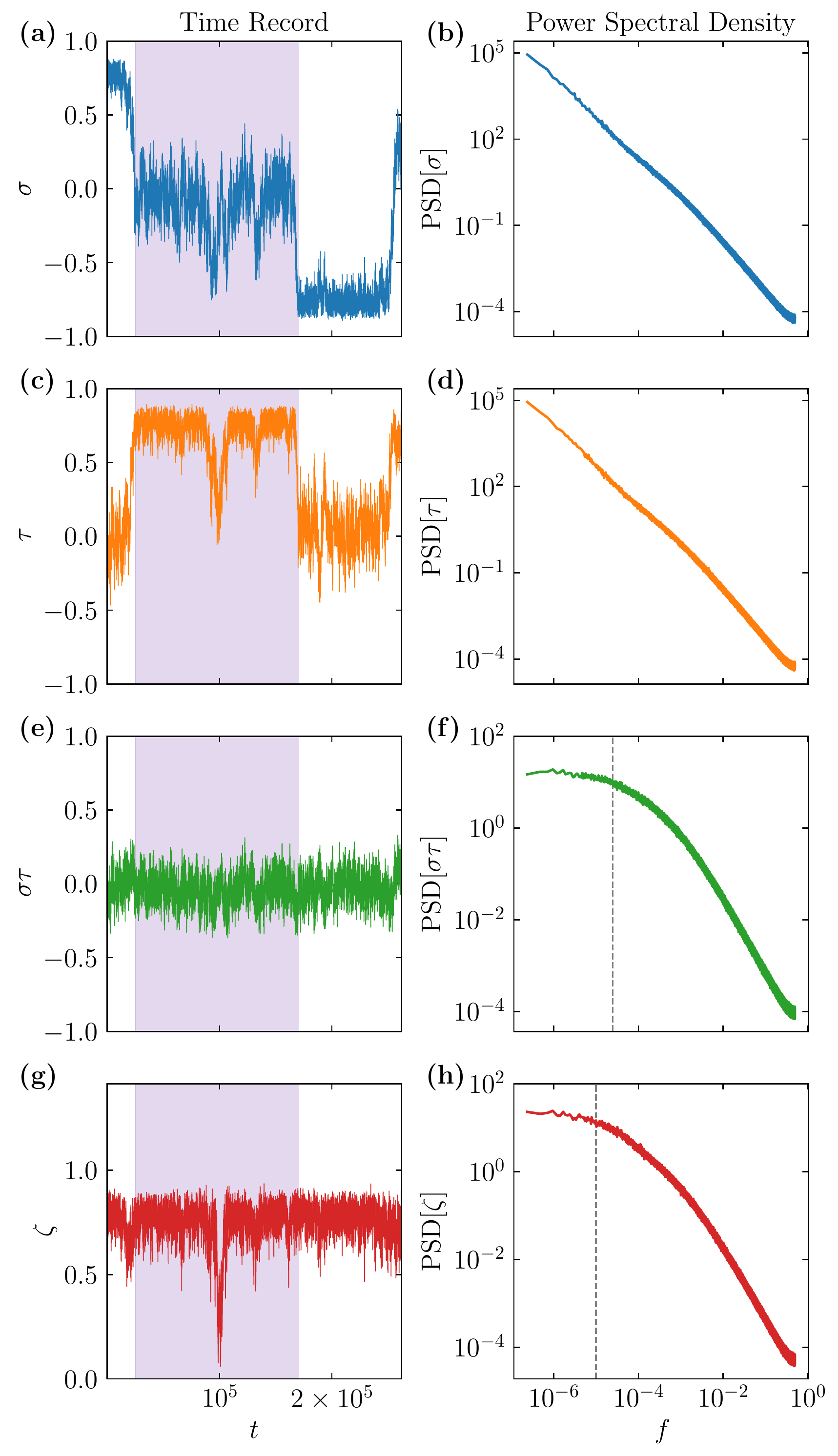}

\caption{Metropolis evolution of the RBFM for a system size $L=80$, $K=0$,
and $\varepsilon=0.5\,J$ at the peak temperature from Fig. \ref{fig:K-0.0_pseudo-Tc_and_peak_value}
$(T_{\chi}=2.129\,J)$. We show a typical time record as a function
of Monte Carlo sweeps $t$ and the disorder-averaged power spectral
density (PSD) as a function of frequency $f$ for the \textbf{(a,
b)} $\sigma$, \textbf{(c, d)} $\tau$, \textbf{(e, f)} Baxter $\sigma\tau$,
and \textbf{(g, h)} quadrature $\zeta$ variables. The vertical lines
correspond approximately to the frequencies at which the PSDs crossover
from a plateau to a power-law behavior, if plateaus are present. The
highlighted region of the time records is discussed in the main text.
\label{fig:Metropolis-evolved-time-records_h-0.5}}
\end{minipage}
\end{figure*}

\subsection{Dynamics at zero Baxter coupling}

The time records for the magnetic, Baxter, and quadrature variables
are shown in Figs. \ref{fig:Metropolis-evolved-time-records_h-0.0}
and \ref{fig:Metropolis-evolved-time-records_h-0.5} for a clean and
a disordered system, respectively. These time records are typical among
different disorder configurations. Additionally, the averaged PSDs
for each observable are shown in the same figures. The time records
are only shown over about 250,000 Monte Carlo sweeps to make the switching
events clearer.

The clean-case time records in Fig. \ref{fig:Metropolis-evolved-time-records_h-0.0}
show multiple-state switching events with the longest timescale being
about 100,000 Monte Carlo sweeps. This is reflected in the averaged
PSDs in the same figure, which exhibit a plateau at a frequency scale
of $10^{-5}\,{\rm sweeps}^{-1}$. Additionally, the switching of the
two magnetic variables appear independent, and their averaged magnetic
PSDs appear identical. This is consistent with the fact that the clean
system with $K=0$ corresponds to two independent Ising models. Since
the Baxter variable $\sigma\tau$ and the quadrature variable fluctuate
whenever either $\sigma$ or $\tau$ fluctuate, the typical timescales
associated with these variables is smaller, with the longest one being
about 50,000 for the Baxter variable. This smaller timescale is seen
in the Baxter PSD as well, as the plateau is shifted by about a factor
of two along the frequency axis. The quadrature variable fluctuates
the fastest when $\sigma$ and $\tau$ are simultaneously near zero.

The corresponding time records and PSDs for the disordered system
in Fig. \ref{fig:Metropolis-evolved-time-records_h-0.5} show that
the Monte Carlo dynamics of the fluctuations are qualitatively different
than they are for the clean case. The most obvious difference is that
the amplitude of the Baxter fluctuations is reduced. Similarly, the
relative change in the quadrature variable is typically smaller than
that in the clean system (with the exception of an event occurring
near 100,000 Monte Carlo sweeps). The reduction of the Baxter noise
can be attributed to domain break-up within the system, whereas that
for the quadrature fluctuations is due to how infrequently both magnetic
variables vanish simultaneously.

The main difference between the clean and disordered cases is that
magnetic fluctuations are correlated in the latter due to the random
Baxter field. Consider, for example, the time records between about
25,000 and 170,000 Monte Carlo sweeps, as highlighted in the purple
region in Fig. \ref{fig:Metropolis-evolved-time-records_h-0.5}. During
this time period, the $\sigma$-magnetic variable is mostly fluctuating
around $\sigma=0$ and the $\tau$-magnetic variable is fluctuating
around $\tau\approx0.75$. But for times preceding this interval,
or those immediately following it, the magnetic variables change roles
with $\sigma$ fluctuating around a finite value and $\tau$ fluctuating
around zero. These relatively long time periods in which one magnetic
variable has a finite expectation value while the other is fluctuating
around zero correspond precisely to the hollow-square distribution
of $\left(\sigma,\tau\right)$ states shown in Fig. \ref{fig:K-0.0_h-0.5_probabilities}
at temperatures $T<1.095\,T_{\chi}$. The event around 100,000 Monte
Carlo sweeps, which substantially changes the quadrature variable,
seems to correspond to a failed, simultaneous switching of $\sigma$
and $\tau$.

The differences in the time records of the clean and disordered cases
are also manifested in the PSDs. In both cases, the Baxter and quadrature
variables show a plateau within the accessible frequency range, indicative
of the longest timescale associated with these fluctuations \citep{MacDonald_D_K_C_2006-06-23}.
However, plateaus for the $\sigma$ and $\tau$ magnetic are absent
in the disordered case, in contrast with the clean system. Indeed,
the magnetic PSDs display power-law behavior down to the smallest
frequencies accessible in the disordered system. This behavior indicates
that the longest timescale associated with the magnetic variables
is too long and inaccessible during these simulations, and at least
two orders of magnitude longer than the Baxter timescale.

At first, one might have expected the Baxter and magnetic timescales
to be comparable, since the Baxter domains fluctuate only because
the magnetic degrees of freedom fluctuate. Indeed, referring to the
purple region in the time records of Fig. \ref{fig:Metropolis-evolved-time-records_h-0.5},
the $\sigma$ variable is fluctuating around zero, $\tau$ remains
relatively constant, and the Baxter variable also fluctuates around
zero. This shows that, overall, the $\sigma$ variable is broken apart
to satisfy the random strain constraint, allowing $\tau$ to take
on a finite value. This would seemingly imply that the $\sigma$ timescale
should be comparable to the Baxter one, while the timescale for $\tau$
should be longer than both.

However, the event around 100,000 sweeps, in which there is a failure
for $\sigma$ and $\tau$ to switch roles, provides the explanation
of why the magnetic timescales can be so much longer than the Baxter
ones. While the Baxter variable fluctuates, the magnetic variables
need to simultaneously break apart and reverse roles for a timescale
to appear in the PSDs. But such events are rare, since the constraints
imposed by the random Baxter field are satisfied by only a single
magnetic variable. The scarcity of these events is what then leads
to the long time magnetic scales and to disappearance of the plateaus
in the magnetic PSDs.

\section{Discussion\label{sec:Discussion}}

The RBFM proposed and studied here provides a simple yet powerful
framework to capture the impact of random strain in nematic systems
for which the nematic instability is intertwined with a magnetic (or
charge) stripe instability. Because in this case the nematic is a
composite order parameter, random strain has a dual role as a random
nematic field and a random magnetic bond. The extensive Monte Carlo
simulations performed here reveal that, like in the RFIM, which is
suitable for systems for which nematicity is a primary instability,
the nematic ground state in the RBFM also breaks up in domains due
to the random strain. However, in the RBFM, random strain also promotes
correlations between the magnetic variables that have no counterpart
either in the RFIM or in the clean Ashkin-Teller model. These correlations
are ultimately a consequence of the fact that the constraints imposed
by the random strain can be satisfied locally by two types of magnetic
configurations, resulting in a residual degeneracy for the system.
This property, in turn, is a direct manifestation of the composite
character of the nematic order parameter.

We showed that these random-strain promoted correlations are manifested
in several properties of the RBFM. They not only promote the emergence
of a second break-up length scale, but also of a second much longer
time scale associated with the simultaneous switching of the two magnetic
order parameters. These correlations appear unambigously in the magnetic
configurational space as a probability distribution of the $\{\sigma,\tau\}$
states in the shape of a hollow square. Such a distribution function
is interpreted in terms of one magnetic Ising variable fluctuating
around zero to satisfy the random-strain constraint, while the other
one acquires a nonzero average value. The main signature of the disorder-promoted
correlations in the thermodynamics is the unexpected enhancement of
the magnetic susceptibility for large enough disorder strength. In
contrast, nematic fluctuations are generally suppressed by disorder,
similarly to the RFIM.

The fact that multiple magnetic configurations can satisfy a given
random-strain realization suggests that, in contrast to the RFIM,
the RBFM may have a macroscopic ground state degeneracy. While our
attempts to probe this macroscopic ground state degeneracy via Replica-Exchange
WL simulations gave inconclusive results, if this turns out to be
the case, a spin glass phase for the magnetic variables might emerge.
Indeed, for the $K=0$ case, the Hamiltonian in Eq. (\ref{eq:RBFATM_Hamiltonian})
has the form of two independent Ising models coupled by a random bond.
This bilayer system is similar to the Edwards-Anderson (EA) model
of spin glasses \citep{EdwardsAnderson_1975}, although in this case,
the two Ising layers representing the two magnetic variables in the
RBFM would have clean Ising ferromagnetic exchanges within each layer,
but the layers would be randomly coupled to each other locally due to random strain.
It would be interesting to see if this EA implementation would also
realize a low-temperature spin glass phase characterized by replica-symmetry
breaking \citep{BinderYoung1986spin_glass_review}.

The main material candidates to observe these effects are the iron
pnictides of the 122 (e.g. BaFe$_{2}$As$_{2}$), 1111 (e.g. LaFeAsO),
and 111 (e.g. NaFeAs) families. In all these cases, nematic order
is intertwined with an antiferromagnetic stripe order, and there is
strong evidence for the magnetic origin of the nematic instability
\citep{fernandes2014what_drives}. In the language of the RBFM, onto
which the Ising $J_{1}$-$J_{2}$ model can be mapped, the $\sigma$
and $\tau$ variables refer to the staggered magnetizations of the
two interpenetrating N\'eel sublattices that form the stripe state whereas
the Baxter variable refers to the composite nematic order parameter (recall
Fig. \ref{fig:J1-J2_ground_states}). In these and other materials,
random strain is promoted by both chemical substitution and intrinsic
lattice defects such as vacancies, interstitials, dislocations, and
twin boundaries \citep{egami_2012_structural_variation_prb}. These
intrinsic effects can be partially remedied via annealing, which was
shown in CaFe$_{2}$As$_{2}$ to reduce the amount of random strain
present in the sample \citep{canfield_2011_annealing_pressure_prb}.
Interestingly, in BaFe$_{2}$As$_{2}$, annealing was found to bring
the magnetic and nematic transitions closer together towards a simultaneous
first-order transition \citep{Birgeneau_2016_post-growth_annealing_iop}.
A similar reduction of the splitting between the two transitions was
seen in $\mathrm{CeFeAsO}$ single crystals when compared to polycristalline samples,
which presumably have larger intrinsic random strain \citep{geibel_2010_CeFeAsO_prb}.
These observations, as well as the general increase of the magnetic
transition temperature in annealed samples \citep{geibel_2010_CeFeAsO_prb,canfield_2011_annealing_pressure_prb,Birgeneau_2016_post-growth_annealing_iop},
are qualitatively consistent with the splitting between the peak temperatures
of the magnetic and Baxter susceptibilities seen in our simulations
(see Fig. \ref{fig:K-0.0_pseudo-Tc_and_peak_value}). Of course, while
in our 2D RBFM model no long-range order is allowed, the magnetic
and nematic transitions are expected to survive up to a finite disorder
strength in a more realistic model with coupled RBFM layers.

Evidence for inhomogeneous and glassy-like fluctuations was reported
by nuclear magnetic resonance (NMR) measurements in 1111 \citep{grafe_2013_nmr_prb_LaFeAsO}
and 122 compounds \citep{curro_2015_nmr_prb,curro_2016_nmr_prl}.
In the case of LaFeAsO, nuclear quadrupole resonance (NQR) measurements
further reported the existence of different local charge environments
\citep{Lang2010}. While it is tempting to associate these behaviors
with nematic and magnetic domain break-up, and to speculate that the
two charge environments could correspond to the two types of domain
break-up, further analysis is needed to disentangle this from other
possible effects. Indeed, because the spin $3/2$ of the $^{75}\mathrm{As}$
nucleus experiences both dipolar and quadrupolar interactions \citep{curro_2016_nmr_prl},
it would be valuable to quantitatively separate the nematic and magnetic
fluctuations contributions to the spin-lattice relaxation rate. Moreover,
a systematic analysis of the impact of annealing on the glassy-like
NMR response would be desirable. It is also interesting to note that
ultrafast optical measurements in the magnetically ordered state of
BaFe$_{2}$As$_{2}$ revealed the existence of two different relaxation
time scales, a ``slow'' one (of the order of tens of picoseconds)
and a ``fast'' one (of the order of picoseconds) \citep{Patz2014}.
While a different interpretation was proposed in that paper, additional
experiments in samples subjected to different annealing regimens could
help elucidate whether these two times scales are those that characterize
the magnetic and nematic degrees of freedom in the RBFM.

We conclude by emphasizing that the RBFM should also describe realistic
nematic phenomena in other quasi-2D quantum materials for which nematicity
is a partially melted density-wave stripe state. Beyond the iron pnictides,
nematicity in the cuprates has been proposed to arise from the partial
melting of charge stripe order \citep{Nie2014}. In contrast to magnetic
domains, but like nematic domains, charge-order domains can in principle
also be probed via scanning tunneling microscopy (STM) \citep{Lawler2010,Mesaros2011}.
This opens up the interesting prospect of a quantitative analysis
to search for signatures of RBFM behavior in the STM data, similarly
to what has been previously done using the RFIM \citep{Phillabaum2012}.
\begin{acknowledgments}
We thank A. Chakraborty, J. Freedberg, and E. D. Dahlberg for fruitful
discussions. WJM and RMF were supported by the U. S. Department of
Energy, Office of Science, Basic Energy Sciences, Materials Sciences
and Engineering Division, under Award No. DE-SC0020045. TV
was supported by the National Science Foundation under Grant No. DMR-1828489.
TV and RMF acknowledge the hospitality of
KITP at UCSB, where the work was initiated. KITP is supported by the
National Science Foundation under Grant No. NSF PHY-1748958. We thank
the Minnesota Supercomputing Institute (MSI) at the University of
Minnesota, where the numerical computations were performed.
\end{acknowledgments}

\appendix

\section{Numerical Methods\label{sec:Numerical-Methods}}

\subsection{Adaptive simulated annealing \label{subsec:simulated_annealing}}

We used simulated annealing to get a qualitative picture of the emergence
of two lengths scales in the RBFM induced by the random Baxter field.
To do so, we had to simulate the system at low temperatures to disentangle
the disorder effects from thermal fluctuations effects. At low temperatures,
the effects of disorder are more pronounced, but this leads to the
possibility of a simulation getting stuck in a metastable state if
one were to simply set the temperature to be low. Simulated annealing
provides a physically-motivated way to move a simulation into lower
energy states by thermally activated tunneling through energy barriers
separating local energy minima \citep{numerical_recipes_simulated_annealing}.
It does so by first raising the temperature of the simulation above
all energy barriers, and then slowly lowering the temperature of the
system, or ``annealing'' it, until the lowest temperature is reached.
Our algorithm is ``adaptive'' in the sense that it continues simulating
at a single temperature if it notices the system is not yet equilibrated.

A conventional simulated annealing regimen is performed according
to the following steps. First, one initializes the system in some
state. Then, an initial temperature is chosen and the system is equilibrated
using some updating scheme satisfying detailed balance. From there,
the system is annealed by lowering the temperature slowly enough such
that it equilibrates at each new temperature until it reaches some
defined lowest temperature \citep{numerical_recipes_simulated_annealing}.

Our adaptive approach works by defining equilibrium based on time
records of the energy that are $L_{t}$ sweeps long. The time records
are split into two blocks and the system is said to be equilibrated
if one of the following two conditions are met:
\begin{enumerate}
\item The energy does not vary in both blocks of Monte Carlo time,
\item The average energy of the second block is within one standard deviation
of the first \emph{and} the variance of the energy in the second block
is within a specified tolerance $\eta$ of that in the first block.
\end{enumerate}
The first condition is that the system is frozen in a particular energy
state -- this may only occur at low temperatures. Condition 2 is
that measurements of the energy and specific heat (through Eq. (\ref{eq:specific_heat}))
are statistically time-independent. If neither of these two conditions
are met, then the system continues its Markov chain at the same temperature
and computes another time record of the energy.

The numerical parameters used in our simulated annealing run are as
follows. We initialized the magnetic degrees of freedom in the RBFM
with equal probability of being $\pm1$ at each site, which corresponds
to a microstate with maximal entropy. We used the initial temperature
of $T_{i}\equiv2.269\:J$ corresponding to the 2D Ising critical temperature.
We equilibrated the system using a single-site Metropolis updating
scheme over two time record blocks, each of size $2^{10}=1024$ sweeps,
making $L_{t}=2048$ sweeps long. The specific heat tolerance was
set to $\eta=0.05$. We chose a cooling regimen corresponding to
\begin{equation}
T_{j}=T_{i}\left(\frac{T_{f}}{T_{i}}\right)^{\nicefrac{j}{N_{{\rm steps}}}},
\end{equation}
where $T_{j}$ is the temperature of the $j^{{\rm th}}$ simulated
annealing step, with $j\in\left\{ 1,2,\dots,N_{{\rm steps}}\right\} $,
and $T_{f}$ is the final temperature. The data shown in Figs. \ref{fig:domain_decomposition_simulated_annealing}
and \ref{fig:Isotropic-spin-spin-correlation_functions} from simulated
annealing had a minimum temperature of $T_{f}=0.269\,J$ and ran over
$N_{{\rm steps}}=150$ steps. Thus, the temperature of the simulation
decreased by about 1.4\% between steps. At this small percentage,
we noticed that each step would typically require less than five adaptive
iterations to equilibrate.

\subsection{Wang Landau sampling}

We employ the Replica-Exchange Wang-Landau (WL) Monte Carlo technique
recently developed for complex energy landscapes \citep{vogel2013REWLgeneric,vogel2014REWLscalable,vogel2018REWLtutorial}
to simulate Eq. (\ref{eq:RBFATM_Hamiltonian}). This technique expands
on the WL sampling algorithm known for its ability to quickly calculate
thermodynamic properties at arbitrary temperatures after obtaining
a system's microcanonical density of states \citep{wang2001efficient_1stpaper,wang2001determining}.

In the WL algorithm, the density of states $g$ at a given energy
$E$ is obtained iteratively by performing a random walk through energy
space. It is calculated by keeping track of two histograms during
a random walk through the system's phase space. The first is the approximation
of $g(E)$ and the other one is typically called the ``energy histogram,''
denoted by $\mathcal{H}(E)$. The energy histogram is used to ensure
the sampling of the phase space is uniform throughout the simulation.

The density of states is initially approximated as a constant of value
one whereas the energy histogram is set to be a constant zero. During
the random walk, transitions from energy $E$ to $E^{\prime}$ are
accepted with probability
\begin{equation}
\mathcal{W}(E\rightarrow E^{\prime})={\rm min}\left\{ 1,\frac{g(E)}{g(E^{\prime})}\right\} .\label{eq:WL_transition_probability}
\end{equation}
 In our simulations, new energy states are generated as single-spin
updates of one Ising spin color for a single lattice sweep while the
other one is quenched. Regardless of whether the transition is accepted,
one updates both $g$ and $\mathcal{H}$ at the resulting energy.
The increment for the energy histogram always comes from the addition
of 1 in the resulting energy bin. The density of states in that energy
bin, meanwhile, is scaled by a factor $f_{i}$ in the $i^{{\rm th}}$
iteration, where $f_{i}$ is given by
\begin{align}
f_{i} & =f_{1}^{\frac{1}{2^{i-1}}}.
\end{align}
 A typical value for the initial increment is $f_{1}=e$. Each iteration
ends once the energy histogram is ``flat.'' In our simulations,
``flatness'' was measured according to a tolerance $\epsilon\leq0.3$
such that
\begin{equation}
\frac{\max\left\{ \mathfrak{\mathcal{H}}(E)\right\} -\min\left\{ \mathfrak{\mathcal{H}}(E)\right\} }{\min\left\{ \mathfrak{\mathcal{H}}(E)\right\} }<\epsilon.
\end{equation}
 After an iteration, the energy histogram is reset to zero and, conventionally,
the density of states is left untouched. However, since $g(E)$ updated
in the WL scheme is only the relative density of states, after each
iteration we scale the density of states by its minimum value to maintain
numerical sensitivity to the decreasing $f_{i}$.

The microcanonical density of states is considered ``converged''
when its increment is smaller than a predefined value. In our case,
this value was $f_{i}\leq1+5\times10^{-7}$ in the most demanding
cases ($K=0$ simulations), although we typically used a more stringent
condition of $f_{i}\leq1+1\times10^{-7}$ when possible ($K>0$ simulations).
Upon convergence, the thermodynamic observables obtained by energetic
moments can be computed from the density of states at any temperature
$T$, since the partition function is given by
\begin{equation}
\mathcal{Z}=\sum_{E}g(E)\,{\rm e}^{-E/T}.\label{eq:partition_function}
\end{equation}
For example, $\left\langle E^{n}\right\rangle $ for some power $n$
is obtained by
\begin{equation}
\left\langle E^{n}\right\rangle =\frac{1}{\mathcal{Z}}\sum_{E}g(E)\,{\rm e}^{-E/T}\,E^{n},
\end{equation}
from which the specific heat, measured as the heat capacity per lattice
site, is calculated from Eq. (\ref{eq:specific_heat}).

One can extend the WL algorithm to a joint density of states if non-energy
moments of the Boltzmann distribution are required \citep{wang2001efficient_1stpaper,wang2001determining}.
This would require a joint energy histogram whose flatness would then
be measured in the higher dimensional space. However, to speed our
simulations, we instead kept track of another histogram for each non-energy
observable of interest; $\left|\sigma\right|$ and $\left|\tau\right|$
are a couple, for example. These observable histograms do not influence
the convergence rate of the simulation as they do not appear in our
flatness metric. Instead, they capture the mean value of non-energy
observables within each energy bin. These means were updated as a
cumulative moving average per energy bin for each WL update. Denoting
such an observable as $A$, and its histogram of mean values in energy
space as $\overline{A}(E)$, then its expectation value at any temperature
is
\begin{equation}
\left\langle A\right\rangle =\frac{1}{\mathcal{Z}}\sum_{E}g(E)\:{\rm e}^{-E/T}\,\overline{A}(E).\label{eq:non_energy_observable}
\end{equation}
 In our typical simulation, there are more than $10^{7}$ individual
measurements taken for each observable in each energy bin. These measurements
per bin likely suffer from autocorrelations, but we obtain error bars
on thermodynamic observables by running simulations in parallel with
different seeds for the random number generators and then averaging
the results.

\subsection{Replica-Exchange Wang Landau sampling}

WL sampling is a fast Monte Carlo method for studying small and simple
systems. When applied to larger systems, or systems with complex energy
landscapes, the time to solution increases rapidly. Thus, to study
the effect of the random strain disorder in Eq. (\ref{eq:RBFATM_Hamiltonian}),
we employ a Replica-Exchange WL sampling algorithm which is a massively-parallel
extension of the WL procedure above \citep{vogel2013REWLgeneric,vogel2014REWLscalable,vogel2018REWLtutorial}.

Its ability to speed up WL sampling comes in three parts. First, it
divides the phase space volume into a set of $N_{w}$ equally-sized,
overlapping energy windows within which $g(E)$ is approximated using
the WL scheme. Second, the energy windows in the overlapping regions
are occasionally allowed to exchange states with a probability:

\begin{multline}
\mathcal{W}\left(\left\{ E_{1},E_{2}\right\} \rightarrow\left\{ E_{2},E_{1}\right\} \right)\\
=\min\left\{ 1,\frac{g_{1}(E_{1})}{g_{1}(E_{2})}\cdot\frac{g_{2}(E_{2})}{g_{2}(E_{1})}\right\} \label{eq:REWL_transition_probability}
\end{multline}
where $E_{1}$ and $E_{2}$ are the energies of the two windows undergoing
exchange and $g_{1}$ and $g_{2}$ are their density of states. From
Eq. (\ref{eq:WL_transition_probability}), one sees that when the
exchange is not guaranteed, Eq. (\ref{eq:REWL_transition_probability})
is simply the product of the WL transition probabilities for the first
energy window to move from $E_{1}$ to $E_{2}$ and for the second
one to move from $E_{2}$ to $E_{1}$.

When an exchange update is accepted, then the two energy windows involved
swap their spin configurations. The density of states and energy histogram
in each window are always incremented after an exchange attempt, regardless
of whether the exchange actually occurred. In our simulations, an
exchange attempt was made between two adjacent energy windows after
each sweep of the lattice to increase the number of exchanges. This
facilitates information dispersal through the divided phase space.

The third part of Replica-Exchange WL sampling comes by adding $n$
independent WL random walkers, or \emph{replicas}, to each of the
$N_{w}$ energy windows. By increasing $n$, the number of exchanges
increases. The optimal speedup of Replica-Exchange WL over standard
WL comes from a combination of the size of the overlap, the number
of energy windows, the number of replicas per window, and how often
the exchange updates are attempted. The optimizing combination of
these parameters is model-dependent, but we found success when allowing
a 75-80\% overlap between energy windows with only 2 ($K>0$) or 1
($K=0$) replicas per window. When $K>0$, the number of windows was
set to be $N_{w}=2L$. However, when $K=0$ with a nonzero disorder
strength, we found that a smaller number of windows allowed for the
fastest convergence with $N_{w}=24$.

In the Replica-Exchange WL scheme, the simulation is finished when
each density of states in each energy window is converged. If some
windows converge before others, then they continue to perform Replica-Exchange
WL updates until the rest of the windows converge as well. This allows
information from the faster-converging regions of phase space to still
diffuse to the slower regions. When there are multiple replicas in
a single energy window, the window itself moves from one iteration
to the next only when every replica's energy histogram is flat. When
this happens, the density of states from each replica is scaled as
for normal WL sampling, and then the densities of states from all
the replicas are averaged and redistributed before the next iteration
is performed. The observables from all the walkers within a single
energy window, however, are only averaged after the simulation is
complete.

Once complete, the density of states over the full phase space is
assembled by stitching together the individual density of states from
each energy window. This procedure is carried out within the overlapping
regions by finding the energy bin where the microcanonical temperatures
$T^{-1}={\rm d}\log g(E)/{\rm d}E$ are the closest between a lower-energy
window and a higher-energy window. The differentiation is performed
with a finite-difference formula accurate up to fourth-order in the
bin size. The observables are averaged within the overlapping regions.
Since multiple windows may be overlapping in the same region, this
may lead to an overwriting of the final density of states. We do not
stop this from happening; however, the observables are still averaged
across all overlapping windows such that statistics from any energy
window with an overwritten density of states is not lost.

\subsection{Disorder averaging\label{subsec:Disorder-averaging_appendix}}

\begin{table}
\begin{centering}
\caption{Number of Replica-Exchange WL simulation disorder configurations for
each set of model parameters for a system size $L=40$. The columns
with $K$ defined correspond to the RBFM while the RFIM column is
for the random field Ising model. These parameters were used in all
figures other than Fig. \ref{fig:c_V-variable_system_size}. \label{tab:Disorder-configurations-averaged}}
\par\end{centering}
\centering{}%
\begin{tabular}{c|c|c|c}
\hline
Disorder Strength & $K=0.5\,J$ & $K=0$ & RFIM\tabularnewline
\hline
$\varepsilon=0.0\,J$ & 64 & 64 & 128\tabularnewline
$\varepsilon=0.2\,J$ & 96 & 96 & 128\tabularnewline
$\varepsilon=0.5\,J$ & 94 & 120 & 126\tabularnewline
$\varepsilon=0.8\,J$ & 246 & 147 & 148\tabularnewline
\hline
\end{tabular}
\end{table}

\begin{table}
\begin{centering}
\caption{Number of Replica-Exchange WL simulation disorder configurations for
each model parameters for the systems in Fig. \ref{fig:c_V-variable_system_size}.
\label{tab:Disorder-configurations-averaged-increasing-size}}
\par\end{centering}
\centering{}%
\begin{tabular}{c|c|c}
\hline
System Size & $\varepsilon=0.0\,J$ & $\varepsilon=0.5\,J$\tabularnewline
\hline
$L=24$ & 64 & 128\tabularnewline
$L=32$ & 64 & 124\tabularnewline
$L=40$ & 64 & 114\tabularnewline
$L=48$ & 64 & 173\tabularnewline
$L=56$ & 64 & 78\tabularnewline
\hline
\end{tabular}
\end{table}

The Hamiltonian in Eq. (\ref{eq:RBFATM_Hamiltonian}) was simulated
for a single field configuration, or single quenched set of random
Baxter field variables $\left\{ \varepsilon_{i}\right\} $, using
the Replica-Exchange WL procedure. Then, the thermodynamics from many
field configurations were averaged to obtain our final results.

Denoting the thermodynamic average of an observable $A$ in disorder configuration
$j$ by $\left\langle A\right\rangle _{j}$, the disorder-averaged expectation value
over $M$ field configurations was calculated as
\begin{align}
\left\llbracket \left\langle A\right\rangle \right\rrbracket  & =\frac{1}{M}\sum_{j=1}^{M}\left\langle A\right\rangle _{j},
\end{align}
where the double brackets $\left\llbracket \cdot\right\rrbracket$ represent the disorder average.
Its standard error of the mean was computed as
\begin{equation}
\delta A_{{\rm d.a.}}=\left[\frac{1}{M\left(M-1\right)}\sum_{j=1}^{M}\left(\left\langle A\right\rangle _{j}-\left\llbracket \left\langle A\right\rangle \right\rrbracket \right)^{2}\right]^{1/2}.
\end{equation}

Tables \ref{tab:Disorder-configurations-averaged} and \ref{tab:Disorder-configurations-averaged-increasing-size}
show the number of distinct disorder configuration averaged over for
the Replica-Exchange WL simulations. The differences in the numbers
of disorder configurations is due to the increasing difficulty to
converge the simulations for large disorder strengths.

\subsection{Replica-Exchange Wang Landau Simulation code details}

The Replica-Exchange WL parallelization was performed processor-wise
on supercomputing nodes at the \emph{Minnesota Supercomputing Institute}.
The Replica-Exchange WL code was benchmarked against the 2D clean
Ising model and 2D clean ATM, as well as a 2D RFIM, all on a periodic
square lattice. The only difference in the code between these test
models and our ATM is the Hamiltonian selected at compile-time. Any
numerical analysis performed after the simulations were complete,
such as disorder-averaging, plotting, \emph{etc.}, was scripted using
Python 3 making use of the \emph{SciPy} libraries. The source code
and  job scripts can be found at Ref. \citep{REWL_Simulator}.

\bibliographystyle{apsrev4-2}
\bibliography{rbfatm_references}

\end{document}